\documentclass{aptpub}
\usepackage{xspace}
\def\babs{\begin{abstract}}             \def\eabs{\end{abstract}}
\def\barr{\begin{array}}                \def\earr{\end{array}}
\def\bcc{\begin{center}}                \def\ecc{\end{center}}

\def\bdes{\begin{description}}          \def\edes{\end{description}}
\def\bdoc{\begin{document}}             \def\edoc{\end{document}}
\def\ben{\begin{enumerate}}             \def\een{\end{enumerate}}
\def\beqn{\begin{eqnarray}}             \def\eeqn{\end{eqnarray}}
\def\beqnl#1{\beqn\label{#1}}           \def\eeqnl#1{\label{#1}\eeqn}
\def\beqnx{\begin{eqnarray*}}           \def\eeqnx{\end{eqnarray*}}
\def\bseqn{\begin{subeqnarray}}         \def\eseqn{\end{subeqnarray}}
\def\beq#1\eeq{\begin{equation}#1\end{equation}}
\def\bal#1\eal{\begin{align}#1\end{align}}
\def\balx#1\ealx{\begin{align*}#1\end{align*}}
\def\beqx{$$}                           \def\eeqx{$$}
\def\bfig{\protect\begin{figure}}       \def\efig{\protect\end{figure}}
\def\bfigx{\protect\begin{figure*}}     \def\efigx{\protect\end{figure*}}
\def\bfl{\begin{flushleft}}             \def\efl{\end{flushleft}}
\def\bfr{\begin{flushright}}            \def\efr{\end{flushright}}
\def\bit{\begin{itemize}}               \def\eit{\end{itemize}}
\def\bpic{\begin{picture}}              \def\epic{\end{picture}}
\def\bqu{\begin{quote}}                 \def\equ{\end{quote}}
\def\bqun{\begin{quotation}}            \def\equn{\end{quotation}}
\def\bsl{\begin{slide}}                 \def\esl{\end{slide}}
\def\btabb{\begin{tabbing}}             \def\etabb{\end{tabbing}}
\def\btabl{\begin{table}}               \def\etabl{\end{table}}
\def\btablx{\begin{table*}}             \def\etablx{\end{table*}}
\def\btab{\begin{tabular}}              \def\etab{\end{tabular}}
\def\btabu{\begin{tabular}}             \def\etabu{\end{tabular}}
\def\btabx{\begin{tabular*}}            \def\etabx{\end{tabular*}}
\def\bbib{\begin{thebibliography}{999}} \def\ebib{\end{thebibliography}}
\def\bver{\begin{verbatim}}             \def\ever{\end{verbatim}}

\def\XS{\xspace}

\def\bm#1{\mbox{\boldmath $#1$}}                

\def\sbm#1{\mbox{\boldmath $#1$}}
\def\sbmp#1{\mbox{\boldmath{$\scriptstyle #1$}}}
\def\sdmp#1{\mbox{\scriptsize #1}}
\def\trans#1{\mbox{$#1^{\sdmp t}$}}
\def\sbv#1{\mbox{\bf #1}}
\def\sbvp#1{\mbox{\scriptsize {\bf #1}}}
\def\sdm#1{\mbox{#1}}

\def\Ab{{\sbm A}\XS}    \def\ab{{\sbm a}\XS}
\def\Bb{{\sbm B}\XS}    \def\bb{{\sbm b}\XS}
\def\Cb{{\sbm C}\XS}    \def\cb{{\sbm c}\XS}
\def\Db{{\sbm D}\XS}    \def\db{{\sbm d}\XS}
\def\Eb{{\sbm E}\XS}    \def\eb{{\sbm e}\XS}
\def\Fb{{\sbm F}\XS}    \def\fb{{\sbm f}\XS}
\def\Gb{{\sbm G}\XS}    \def\gb{{\sbm g}\XS}
\def\Hb{{\sbm H}\XS}    \def\hb{{\sbm h}\XS}
\def\Ib{{\sbm I}\XS}    \def\ib{{\sbm i}\XS}
\def\Jb{{\sbm J}\XS}    \def\jb{{\sbm j}\XS}
\def\Kb{{\sbm K}\XS}    \def\kb{{\sbm k}\XS}
\def\Lb{{\sbm L}\XS}    \def\lb{{\sbm l}\XS}
\def\Mb{{\sbm M}\XS}    \def\mb{{\sbm m}\XS}
\def\Nb{{\sbm N}\XS}    \def\nb{{\sbm n}\XS}
\def\Ob{{\sbm O}\XS}    \def\ob{{\sbm o}\XS}
\def\Pb{{\sbm P}\XS}    \def\pb{{\sbm p}\XS}
\def\Qb{{\sbm Q}\XS}    \def\qb{{\sbm q}\XS}
\def\Rb{{\sbm R}\XS}    \def\rb{{\sbm r}\XS}
\def\Sb{{\sbm S}\XS}    \def\sb{{\sbm s}\XS}
\def\Tb{{\sbm T}\XS}    \def\tb{{\sbm t}\XS}
\def\Ub{{\sbm U}\XS}    \def\ub{{\sbm u}\XS}
\def\Vb{{\sbm V}\XS}    \def\vb{{\sbm v}\XS}
\def\Wb{{\sbm W}\XS}    \def\wb{{\sbm w}\XS}
\def\Xb{{\sbm X}\XS}    \def\xb{{\sbm x}\XS}
\def\Yb{{\sbm Y}\XS}    \def\yb{{\sbm y}\XS}
\def\Zb{{\sbm Z}\XS}    \def\zb{{\sbm z}\XS}

\def\Abt{\trans{\sbm A}\XS}	\def\abt{\trans{\sbm a}\XS}
\def\Bbt{\trans{\sbm B}\XS}	\def\bbt{\trans{\sbm b}\XS}
\def\Cbt{\trans{\sbm C}\XS}	\def\cbt{\trans{\sbm c}\XS}
\def\Dbt{\trans{\sbm D}\XS}	\def\dbt{\trans{\sbm d}\XS}
\def\Ebt{\trans{\sbm E}\XS}	\def\ebt{\trans{\sbm e}\XS}
\def\Fbt{\trans{\sbm F}\XS}	\def\fbt{\trans{\sbm f}\XS}
\def\Gbt{\trans{\sbm G}\XS}	\def\gbt{\trans{\sbm g}\XS}
\def\Hbt{\trans{\sbm H}\XS}	\def\hbt{\trans{\sbm h}\XS}
\def\Ibt{\trans{\sbm I}\XS}	\def\ibt{\trans{\sbm i}\XS}
\def\Jbt{\trans{\sbm J}\XS}	\def\jbt{\trans{\sbm j}\XS}
\def\Kbt{\trans{\sbm K}\XS}	\def\kbt{\trans{\sbm k}\XS}
\def\Lbt{\trans{\sbm L}\XS}	\def\lbt{\trans{\sbm l}\XS}
\def\Mbt{\trans{\sbm M}\XS}	\def\mbt{\trans{\sbm m}\XS}
\def\Nbt{\trans{\sbm N}\XS}	\def\nbt{\trans{\sbm n}\XS}
\def\Obt{\trans{\sbm O}\XS}	\def\obt{\trans{\sbm o}\XS}
\def\Pbt{\trans{\sbm P}\XS}	\def\pbt{\trans{\sbm p}\XS}
\def\Qbt{\trans{\sbm Q}\XS}	\def\qbt{\trans{\sbm q}\XS}
\def\Rbt{\trans{\sbm R}\XS}	\def\rbt{\trans{\sbm r}\XS}
\def\Sbt{\trans{\sbm S}\XS}	\def\sbt{\trans{\sbm s}\XS}
\def\Tbt{\trans{\sbm T}\XS}	\def\tbt{\trans{\sbm t}\XS}
\def\Ubt{\trans{\sbm U}\XS}	\def\ubt{\trans{\sbm u}\XS}
\def\Vbt{\trans{\sbm V}\XS}	\def\vbt{\trans{\sbm v}\XS}
\def\Wbt{\trans{\sbm W}\XS}	\def\wbt{\trans{\sbm w}\XS}
\def\Xbt{\trans{\sbm X}\XS}	\def\xbt{\trans{\sbm x}\XS}
\def\Ybt{\trans{\sbm Y}\XS}	\def\ybt{\trans{\sbm y}\XS}
\def\Zbt{\trans{\sbm Z}\XS}	\def\zbt{\trans{\sbm z}\XS}

\def\Abp{{\sbmp A}\XS}		\def\abp{{\sbmp a}\XS}
\def\Bbp{{\sbmp B}\XS}		\def\bbp{{\sbmp b}\XS}
\def\Cbp{{\sbmp C}\XS}		\def\cbp{{\sbmp c}\XS}
\def\Dbp{{\sbmp D}\XS}		\def\dbp{{\sbmp d}\XS}
\def\Ebp{{\sbmp E}\XS}		\def\ebp{{\sbmp e}\XS}
\def\Fbp{{\sbmp F}\XS}		\def\fbp{{\sbmp f}\XS}
\def\Gbp{{\sbmp G}\XS}		\def\gbp{{\sbmp g}\XS}
\def\Hbp{{\sbmp H}\XS}		\def\hbp{{\sbmp h}\XS}
\def\Ibp{{\sbmp I}\XS}		\def\ibp{{\sbmp i}\XS}
\def\Jbp{{\sbmp J}\XS}		\def\jbp{{\sbmp j}\XS}
\def\Kbp{{\sbmp K}\XS}		\def\kbp{{\sbmp k}\XS}
\def\Lbp{{\sbmp L}\XS}		\def\lbp{{\sbmp l}\XS}
\def\Mbp{{\sbmp M}\XS}		\def\mbp{{\sbmp m}\XS}
\def\Nbp{{\sbmp N}\XS}		\def\nbp{{\sbmp n}\XS}
\def\Obp{{\sbmp O}\XS}		\def\obp{{\sbmp o}\XS}
\def\Pbp{{\sbmp P}\XS}		\def\pbp{{\sbmp p}\XS}
\def\Qbp{{\sbmp Q}\XS}		\def\qbp{{\sbmp q}\XS}
\def\Rbp{{\sbmp R}\XS}		\def\rbp{{\sbmp r}\XS}
\def\Sbp{{\sbmp S}\XS}		\def\sbp{{\sbmp s}\XS}
\def\Tbp{{\sbmp T}\XS}		\def\tbp{{\sbmp t}\XS}
\def\Ubp{{\sbmp U}\XS}		\def\ubp{{\sbmp u}\XS}
\def\Vbp{{\sbmp V}\XS}		\def\vbp{{\sbmp v}\XS}
\def\Wbp{{\sbmp W}\XS}		\def\wbp{{\sbmp w}\XS}
\def\Xbp{{\sbmp X}\XS}		\def\xbp{{\sbmp x}\XS}
\def\Ybp{{\sbmp Y}\XS}		\def\ybp{{\sbmp y}\XS}
\def\Zbp{{\sbmp Z}\XS}		\def\zbp{{\sbmp z}\XS}

\def\Ac{\mbox{$\cal A$}\XS}	%\def\ac{\mbox{$\cal a$}\XS}
\def\Bc{\mbox{$\cal B$}\XS}	%\def\bc{\mbox{$\cal b$}\XS}
\def\Cc{\mbox{$\cal C$}\XS}	%\def\cc{\mbox{$\cal c$}\XS}
\def\Dc{\mbox{$\cal D$}\XS}	%\def\dc{\mbox{$\cal d$}\XS}
\def\Ec{\mbox{$\cal E$}\XS}	%\def\ec{\mbox{$\cal e$}\XS}
\def\Fc{\mbox{$\cal F$}\XS}	%\def\fc{\mbox{$\cal f$}\XS}
\def\Gc{\mbox{$\cal G$}\XS}	%\def\gc{\mbox{$\cal g$}\XS}
\def\Hc{\mbox{$\cal H$}\XS}	%\def\hc{\mbox{$\cal h$}\XS}
\def\Ic{\mbox{$\cal I$}\XS}	%\def\ic{\mbox{$\cal i$}\XS}
\def\Jc{\mbox{$\cal J$}\XS}	%\def\jc{\mbox{$\cal j$}\XS}
\def\Kc{\mbox{$\cal K$}\XS}	%\def\kc{\mbox{$\cal k$}\XS}
\def\Lc{\mbox{$\cal L$}\XS}	%\def\lc{\mbox{$\cal l$}\XS}
\def\Mc{\mbox{$\cal M$}\XS}	%\def\mc{\mbox{$\cal m$}\XS}
\def\Nc{\mbox{$\cal N$}\XS}	%\def\nc{\mbox{$\cal n$}\XS}
\def\Oc{\mbox{$\cal O$}\XS}	%\def\oc{\mbox{$\cal o$}\XS}
\def\Pc{\mbox{$\cal P$}\XS}	%\def\pc{\mbox{$\cal p$}\XS}
\def\Qc{\mbox{$\cal Q$}\XS}	%\def\qc{\mbox{$\cal q$}\XS}
\def\Rc{\mbox{$\cal R$}\XS}	%\def\rc{\mbox{$\cal r$}\XS}
\def\Sc{\mbox{$\cal S$}\XS}	%\def\sc{\mbox{$\cal s$}\XS}
\def\Tc{\mbox{$\cal T$}\XS}	%\def\tc{\mbox{$\cal t$}\XS}
\def\Uc{\mbox{$\cal U$}\XS}	%\def\uc{\mbox{$\cal u$}\XS}
\def\Vc{\mbox{$\cal V$}\XS}	%\def\vc{\mbox{$\cal v$}\XS}
\def\Wc{\mbox{$\cal W$}\XS}	%\def\wc{\mbox{$\cal w$}\XS}
\def\Xc{\mbox{$\cal X$}\XS}	%\def\xc{\mbox{$\cal x$}\XS}
\def\Yc{\mbox{$\cal Y$}\XS}	%\def\yc{\mbox{$\cal y$}\XS}
\def\Zc{\mbox{$\cal Z$}\XS}	%\def\zc{\mbox{$\cal z$}\XS}

\def\Acp{\mbox{$\scriptstyle \cal A$}\XS}
\def\Bcp{\mbox{$\scriptstyle \cal B$}\XS}
\def\Ccp{\mbox{$\scriptstyle \cal C$}\XS}
\def\Dcp{\mbox{$\scriptstyle \cal D$}\XS}
\def\Ecp{\mbox{$\scriptstyle \cal E$}\XS}
\def\Fcp{\mbox{$\scriptstyle \cal F$}\XS}
\def\Gcp{\mbox{$\scriptstyle \cal G$}\XS}
\def\Hcp{\mbox{$\scriptstyle \cal H$}\XS}
\def\Icp{\mbox{$\scriptstyle \cal I$}\XS}
\def\Jcp{\mbox{$\scriptstyle \cal J$}\XS}
\def\Kcp{\mbox{$\scriptstyle \cal K$}\XS}
\def\Lcp{\mbox{$\scriptstyle \cal L$}\XS}
\def\Mcp{\mbox{$\scriptstyle \cal M$}\XS}
\def\Ncp{\mbox{$\scriptstyle \cal N$}\XS}
\def\Ocp{\mbox{$\scriptstyle \cal O$}\XS}
\def\Pcp{\mbox{$\scriptstyle \cal P$}\XS}
\def\Qcp{\mbox{$\scriptstyle \cal Q$}\XS}
\def\Rcp{\mbox{$\scriptstyle \cal R$}\XS}
\def\Scp{\mbox{$\scriptstyle \cal S$}\XS}
\def\Tcp{\mbox{$\scriptstyle \cal T$}\XS}
\def\Ucp{\mbox{$\scriptstyle \cal U$}\XS}
\def\Vcp{\mbox{$\scriptstyle \cal V$}\XS}
\def\Wcp{\mbox{$\scriptstyle \cal W$}\XS}
\def\Xcp{\mbox{$\scriptstyle \cal X$}\XS}
\def\Ycp{\mbox{$\scriptstyle \cal Y$}\XS}
\def\Zcp{\mbox{$\scriptstyle \cal Z$}\XS}

\def\Act{\trans{\cal A}\XS}
\def\Bct{\trans{\cal B}\XS}
\def\Cct{\trans{\cal C}\XS}
\def\Dct{\trans{\cal D}\XS}
\def\Ect{\trans{\cal E}\XS}
\def\Fct{\trans{\cal F}\XS}
\def\Gct{\trans{\cal G}\XS}
\def\Hct{\trans{\cal H}\XS}
\def\Ict{\trans{\cal I}\XS}
\def\Jct{\trans{\cal J}\XS}
\def\Kct{\trans{\cal K}\XS}
\def\Lct{\trans{\cal L}\XS}
\def\Mct{\trans{\cal M}\XS}
\def\Nct{\trans{\cal N}\XS}
\def\Oct{\trans{\cal O}\XS}
\def\Pct{\trans{\cal P}\XS}
\def\Qct{\trans{\cal Q}\XS}
\def\Rct{\trans{\cal R}\XS}
\def\Sct{\trans{\cal S}\XS}
\def\Tct{\trans{\cal T}\XS}
\def\Uct{\trans{\cal U}\XS}
\def\Vct{\trans{\cal V}\XS}
\def\Wct{\trans{\cal W}\XS}
\def\Xct{\trans{\cal X}\XS}
\def\Yct{\trans{\cal Y}\XS}
\def\Zct{\trans{\cal Z}\XS}

\def\Acb{\sbm{\cal A}\XS}
\def\Bcb{\sbm{\cal B}\XS}
\def\Ccb{\sbm{\cal C}\XS}
\def\Dcb{\sbm{\cal D}\XS}
\def\Ecb{\sbm{\cal E}\XS}
\def\Fcb{\sbm{\cal F}\XS}
\def\Gcb{\sbm{\cal G}\XS}
\def\Hcb{\sbm{\cal H}\XS}
\def\Icb{\sbm{\cal I}\XS}
\def\Jcb{\sbm{\cal J}\XS}
\def\Kcb{\sbm{\cal K}\XS}
\def\Lcb{\sbm{\cal L}\XS}
\def\Mcb{\sbm{\cal M}\XS}
\def\Ncb{\sbm{\cal N}\XS}
\def\Ocb{\sbm{\cal O}\XS}
\def\Pcb{\sbm{\cal P}\XS}
\def\Qcb{\sbm{\cal Q}\XS}
\def\Rcb{\sbm{\cal R}\XS}
\def\Scb{\sbm{\cal S}\XS}
\def\Tcb{\sbm{\cal T}\XS}
\def\Ucb{\sbm{\cal U}\XS}
\def\Vcb{\sbm{\cal V}\XS}
\def\Wcb{\sbm{\cal W}\XS}
\def\Xcb{\sbm{\cal X}\XS}
\def\Ycb{\sbm{\cal Y}\XS}
\def\Zcb{\sbm{\cal Z}\XS}

\def\Acbp{\sbmp{\cal A}\XS}
\def\Bcbp{\sbmp{\cal B}\XS}
\def\Ccbp{\sbmp{\cal C}\XS}
\def\Dcbp{\sbmp{\cal D}\XS}
\def\Ecbp{\sbmp{\cal E}\XS}
\def\Fcbp{\sbmp{\cal F}\XS}
\def\Gcbp{\sbmp{\cal G}\XS}
\def\Hcbp{\sbmp{\cal H}\XS}
\def\Icbp{\sbmp{\cal I}\XS}
\def\Jcbp{\sbmp{\cal J}\XS}
\def\Kcbp{\sbmp{\cal K}\XS}
\def\Lcbp{\sbmp{\cal L}\XS}
\def\Mcbp{\sbmp{\cal M}\XS}
\def\Ncbp{\sbmp{\cal N}\XS}
\def\Ocbp{\sbmp{\cal O}\XS}
\def\Pcbp{\sbmp{\cal P}\XS}
\def\Qcbp{\sbmp{\cal Q}\XS}
\def\Rcbp{\sbmp{\cal R}\XS}
\def\Scbp{\sbmp{\cal S}\XS}
\def\Tcbp{\sbmp{\cal T}\XS}
\def\Ucbp{\sbmp{\cal U}\XS}
\def\Vcbp{\sbmp{\cal V}\XS}
\def\Wcbp{\sbmp{\cal W}\XS}
\def\Xcbp{\sbmp{\cal X}\XS}
\def\Ycbp{\sbmp{\cal Y}\XS}
\def\Zcbp{\sbmp{\cal Z}\XS}

\def\Acbt{\trans{\sbm{\cal A}}\XS}
\def\Bcbt{\trans{\sbm{\cal B}}\XS}
\def\Ccbt{\trans{\sbm{\cal C}}\XS}
\def\Dcbt{\trans{\sbm{\cal D}}\XS}
\def\Ecbt{\trans{\sbm{\cal E}}\XS}
\def\Fcbt{\trans{\sbm{\cal F}}\XS}
\def\Gcbt{\trans{\sbm{\cal G}}\XS}
\def\Hcbt{\trans{\sbm{\cal H}}\XS}
\def\Icbt{\trans{\sbm{\cal I}}\XS}
\def\Jcbt{\trans{\sbm{\cal J}}\XS}
\def\Kcbt{\trans{\sbm{\cal K}}\XS}
\def\Lcbt{\trans{\sbm{\cal L}}\XS}
\def\Mcbt{\trans{\sbm{\cal M}}\XS}
\def\Ncbt{\trans{\sbm{\cal N}}\XS}
\def\Ocbt{\trans{\sbm{\cal O}}\XS}
\def\Pcbt{\trans{\sbm{\cal P}}\XS}
\def\Qcbt{\trans{\sbm{\cal Q}}\XS}
\def\Rcbt{\trans{\sbm{\cal R}}\XS}
\def\Scbt{\trans{\sbm{\cal S}}\XS}
\def\Tcbt{\trans{\sbm{\cal T}}\XS}
\def\Ucbt{\trans{\sbm{\cal U}}\XS}
\def\Vcbt{\trans{\sbm{\cal V}}\XS}
\def\Wcbt{\trans{\sbm{\cal W}}\XS}
\def\Xcbt{\trans{\sbm{\cal X}}\XS}
\def\Ycbt{\trans{\sbm{\cal Y}}\XS}
\def\Zcbt{\trans{\sbm{\cal Z}}\XS}

\def\AD{{\sdm A}\XS}
\def\BD{{\sdm B}\XS}
\def\CD{{\sdm C}\XS}
\def\DD{{\sdm D}\XS}
\def\ED{{\sdm E}\XS}
\def\FD{{\sdm F}\XS}
\def\GD{{\sdm G}\XS}
\def\HD{{\sdm H}\XS}
\def\ID{{\sdm I}\XS}
\def\JD{{\sdm J}\XS}
\def\KD{{\sdm K}\XS}
\def\LD{{\sdm L}\XS}
\def\MD{{\sdm M}\XS}
\def\ND{{\sdm N}\XS}
\def\OD{{\sdm O}\XS}
\def\PD{{\sdm P}\XS}
\def\QD{{\sdm Q}\XS}
\def\RD{{\sdm R}\XS}
\def\SD{{\sdm S}\XS}
\def\TD{{\sdm T}\XS}
\def\UD{{\sdm U}\XS}
\def\VD{{\sdm V}\XS}
\def\WD{{\sdm W}\XS}
\def\XD{{\sdm X}\XS}
\def\YD{{\sdm Y}\XS}
\def\ZD{{\sdm Z}\XS}

\def\ADp{{\sdmp A}\XS}
\def\BDp{{\sdmp B}\XS}
\def\CDp{{\sdmp C}\XS}
\def\DDp{{\sdmp D}\XS}
\def\EDp{{\sdmp E}\XS}
\def\FDp{{\sdmp F}\XS}
\def\GDp{{\sdmp G}\XS}
\def\HDp{{\sdmp H}\XS}
\def\IDp{{\sdmp I}\XS}
\def\JDp{{\sdmp J}\XS}
\def\KDp{{\sdmp K}\XS}
\def\LDp{{\sdmp L}\XS}
\def\MDp{{\sdmp M}\XS}
\def\NDp{{\sdmp N}\XS}
\def\ODp{{\sdmp O}\XS}
\def\PDp{{\sdmp P}\XS}
\def\QDp{{\sdmp Q}\XS}
\def\RDp{{\sdmp R}\XS}
\def\SDp{{\sdmp S}\XS}
\def\TDp{{\sdmp T}\XS}
\def\UDp{{\sdmp U}\XS}
\def\VDp{{\sdmp V}\XS}
\def\WDp{{\sdmp W}\XS}
\def\XDp{{\sdmp X}\XS}
\def\YDp{{\sdmp Y}\XS}
\def\ZDp{{\sdmp Z}\XS}

\def\ADt{\trans{\sdm A}\XS} 
\def\BDt{\trans{\sdm B}\XS}
\def\CDt{\trans{\sdm C}\XS}
\def\DDt{\trans{\sdm D}\XS}
\def\EDt{\trans{\sdm E}\XS}
\def\FDt{\trans{\sdm F}\XS}
\def\GDt{\trans{\sdm G}\XS}
\def\HDt{\trans{\sdm H}\XS}
\def\IDt{\trans{\sdm I}\XS}
\def\JDt{\trans{\sdm J}\XS}
\def\KDt{\trans{\sdm K}\XS}
\def\LDt{\trans{\sdm L}\XS}
\def\MDt{\trans{\sdm M}\XS}
\def\NDt{\trans{\sdm N}\XS}
\def\ODt{\trans{\sdm O}\XS}
\def\PDt{\trans{\sdm P}\XS}
\def\QDt{\trans{\sdm Q}\XS}
\def\RDt{\trans{\sdm R}\XS}
\def\SDt{\trans{\sdm S}\XS}
\def\TDt{\trans{\sdm T}\XS}
\def\UDt{\trans{\sdm U}\XS}
\def\VDt{\trans{\sdm V}\XS}
\def\WDt{\trans{\sdm W}\XS}
\def\XDt{\trans{\sdm X}\XS}
\def\YDt{\trans{\sdm Y}\XS}
\def\ZDt{\trans{\sdm Z}\XS}

\def\aD{{\sdm a}\XS}
\def\bD{{\sdm b}\XS}
\def\cD{{\sdm c}\XS}
\def\dD{{\sdm d}\XS}
\def\eD{{\sdm e}\XS}
\def\fD{{\sdm f}\XS}
\def\gD{{\sdm g}\XS}
\def\hD{{\sdm h}\XS}
\def\iD{{\sdm i}\XS}
\def\jD{{\sdm j}\XS}
\def\kD{{\sdm k}\XS}
\def\lD{{\sdm l}\XS}
\def\mD{{\sdm m}\XS}
\def\nD{{\sdm n}\XS}
\def\oD{{\sdm o}\XS}
\def\pD{{\sdm p}\XS}
\def\qD{{\sdm q}\XS}
\def\rD{{\sdm r}\XS}
\def\sD{{\sdm s}\XS}
\def\tD{{\sdm t}\XS}
\def\uD{{\sdm u}\XS}
\def\vD{{\sdm D}\XS}
\def\wD{{\sdm w}\XS}
\def\xD{{\sdm x}\XS}
\def\yD{{\sdm y}\XS}
\def\zD{{\sdm z}\XS}

\def\aDp{{\sdmp a}\XS}
\def\bDp{{\sdmp b}\XS}
\def\cDp{{\sdmp c}\XS}
\def\dDp{{\sdmp d}\XS}
\def\eDp{{\sdmp e}\XS}
\def\fDp{{\sdmp f}\XS}
\def\gDp{{\sdmp g}\XS}
\def\hDp{{\sdmp h}\XS}
\def\iDp{{\sdmp i}\XS}
\def\jDp{{\sdmp j}\XS}
\def\kDp{{\sdmp k}\XS}
\def\lDp{{\sdmp l}\XS}
\def\mDp{{\sdmp m}\XS}
\def\nDp{{\sdmp n}\XS}
\def\oDp{{\sdmp o}\XS}
\def\pDp{{\sdmp p}\XS}
\def\qDp{{\sdmp q}\XS}
\def\rDp{{\sdmp r}\XS}
\def\sDp{{\sdmp s}\XS}
\def\tDp{{\sdmp t}\XS}
\def\uDp{{\sdmp u}\XS}
\def\vDp{{\sdmp v}\XS}
\def\wDp{{\sdmp w}\XS}
\def\xDp{{\sdmp x}\XS}
\def\yDp{{\sdmp y}\XS}
\def\zDp{{\sdmp z}\XS}

\def\aDt{\trans{\sdm a}\XS}
\def\bDt{\trans{\sdm b}\XS}
\def\cDt{\trans{\sdm c}\XS}
\def\dDt{\trans{\sdm d}\XS}
\def\eDt{\trans{\sdm e}\XS}
\def\fDt{\trans{\sdm f}\XS}
\def\gDt{\trans{\sdm g}\XS}
\def\hDt{\trans{\sdm h}\XS}
\def\iDt{\trans{\sdm i}\XS}
\def\jDt{\trans{\sdm j}\XS}
\def\kDt{\trans{\sdm k}\XS}
\def\lDt{\trans{\sdm l}\XS}
\def\mDt{\trans{\sdm m}\XS}
\def\nDt{\trans{\sdm n}\XS}
\def\oDt{\trans{\sdm o}\XS}
\def\pDt{\trans{\sdm p}\XS}
\def\qDt{\trans{\sdm q}\XS}
\def\rDt{\trans{\sdm r}\XS}
\def\sDt{\trans{\sdm s}\XS}
\def\tDt{\trans{\sdm t}\XS}
\def\uDt{\trans{\sdm u}\XS}
\def\vDt{\trans{\sdm v}\XS}
\def\wDt{\trans{\sdm w}\XS}
\def\xDt{\trans{\sdm x}\XS}
\def\yDt{\trans{\sdm y}\XS}
\def\zDt{\trans{\sdm z}\XS}

\def\Av{{\sbv A}\XS}	\def\av{{\sbv a}\XS}
\def\Bv{{\sbv B}\XS}	\def\bv{{\sbv b}\XS}
\def\Cv{{\sbv C}\XS}	\def\cv{{\sbv c}\XS}
\def\Dv{{\sbv D}\XS}	\def\dv{{\sbv d}\XS}
\def\Ev{{\sbv E}\XS}	\def\ev{{\sbv e}\XS}
\def\Fv{{\sbv F}\XS}	\def\fv{{\sbv f}\XS}
\def\Gv{{\sbv G}\XS}	\def\gv{{\sbv g}\XS}
\def\Hv{{\sbv H}\XS}	\def\hv{{\sbv h}\XS}
\def\Iv{{\sbv I}\XS}	\def\iv{{\sbv i}\XS}
\def\Jv{{\sbv J}\XS}	\def\jv{{\sbv j}\XS}
\def\Kv{{\sbv K}\XS}	\def\kv{{\sbv k}\XS}
\def\Lv{{\sbv L}\XS}	\def\lv{{\sbv l}\XS}
\def\Mv{{\sbv M}\XS}	\def\mv{{\sbv m}\XS}
\def\Nv{{\sbv N}\XS}	\def\nv{{\sbv n}\XS}
\def\Ov{{\sbv O}\XS}	\def\ov{{\sbv o}\XS}
\def\Pv{{\sbv P}\XS}	\def\pv{{\sbv p}\XS}
\def\Qv{{\sbv Q}\XS}	\def\qv{{\sbv q}\XS}
\def\Rv{{\sbv R}\XS}	\def\rv{{\sbv r}\XS}
\def\Sv{{\sbv S}\XS}	\def\sv{{\sbv s}\XS}
\def\Tv{{\sbv T}\XS}	\def\tv{{\sbv t}\XS}
\def\Uv{{\sbv U}\XS}	\def\uv{{\sbv u}\XS}
\def\Vv{{\sbv V}\XS}	\def\vv{{\sbv v}\XS}
\def\Wv{{\sbv W}\XS}	\def\wv{{\sbv w}\XS}
\def\Xv{{\sbv X}\XS}	\def\xv{{\sbv x}\XS}
\def\Yv{{\sbv Y}\XS}	\def\yv{{\sbv y}\XS}
\def\Zv{{\sbv Z}\XS}	\def\zv{{\sbv z}\XS}

\def\Avp{{\sbvp A}\XS}	\def\avp{{\sbvp a}\XS}
\def\Bvp{{\sbvp B}\XS}	\def\bvp{{\sbvp b}\XS}
\def\Cvp{{\sbvp C}\XS}	\def\cvp{{\sbvp c}\XS}
\def\Dvp{{\sbvp D}\XS}	\def\dvp{{\sbvp d}\XS}
\def\Evp{{\sbvp E}\XS}	\def\evp{{\sbvp e}\XS}
\def\Fvp{{\sbvp F}\XS}	\def\fvp{{\sbvp f}\XS}
\def\Gvp{{\sbvp G}\XS}	\def\gvp{{\sbvp g}\XS}
\def\Hvp{{\sbvp H}\XS}	\def\hvp{{\sbvp h}\XS}
\def\Ivp{{\sbvp I}\XS}	\def\ivp{{\sbvp i}\XS}
\def\Jvp{{\sbvp J}\XS}	\def\jvp{{\sbvp j}\XS}
\def\Kvp{{\sbvp K}\XS}	\def\kvp{{\sbvp k}\XS}
\def\Lvp{{\sbvp L}\XS}	\def\lvp{{\sbvp l}\XS}
\def\Mvp{{\sbvp M}\XS}	\def\mvp{{\sbvp m}\XS}
\def\Nvp{{\sbvp N}\XS}	\def\nvp{{\sbvp n}\XS}
\def\Ovp{{\sbvp O}\XS}	\def\ovp{{\sbvp o}\XS}
\def\Pvp{{\sbvp P}\XS}	\def\pvp{{\sbvp p}\XS}
\def\Qvp{{\sbvp Q}\XS}	\def\qvp{{\sbvp q}\XS}
\def\Rvp{{\sbvp R}\XS}	\def\rvp{{\sbvp r}\XS}
\def\Svp{{\sbvp S}\XS}	\def\svp{{\sbvp s}\XS}
\def\Tvp{{\sbvp T}\XS}	\def\tvp{{\sbvp t}\XS}
\def\Uvp{{\sbvp U}\XS}	\def\uvp{{\sbvp u}\XS}
\def\Vvp{{\sbvp V}\XS}	\def\vvp{{\sbvp v}\XS}
\def\Wvp{{\sbvp W}\XS}	\def\wvp{{\sbvp w}\XS}
\def\Xvp{{\sbvp X}\XS}	\def\xvp{{\sbvp x}\XS}
\def\Yvp{{\sbvp Y}\XS}	\def\yvp{{\sbvp y}\XS}
\def\Zvp{{\sbvp Z}\XS}	\def\zvp{{\sbvp z}\XS}

\def\Avt{\trans{\sbv A}\XS}	\def\avt{\trans{\sbv a}\XS}
\def\Bvt{\trans{\sbv B}\XS}	\def\bvt{\trans{\sbv b}\XS}
\def\Cvt{\trans{\sbv C}\XS}	\def\cvt{\trans{\sbv c}\XS}
\def\Dvt{\trans{\sbv D}\XS}	\def\dvt{\trans{\sbv d}\XS}
\def\Evt{\trans{\sbv E}\XS}	\def\evt{\trans{\sbv e}\XS}
\def\Fvt{\trans{\sbv F}\XS}	\def\fvt{\trans{\sbv f}\XS}
\def\Gvt{\trans{\sbv G}\XS}	\def\gvt{\trans{\sbv g}\XS}
\def\Hvt{\trans{\sbv H}\XS}	\def\hvt{\trans{\sbv h}\XS}
\def\Ivt{\trans{\sbv I}\XS}	\def\ivt{\trans{\sbv i}\XS}
\def\Jvt{\trans{\sbv J}\XS}	\def\jvt{\trans{\sbv j}\XS}
\def\Kvt{\trans{\sbv K}\XS}	\def\kvt{\trans{\sbv k}\XS}
\def\Lvt{\trans{\sbv L}\XS}	\def\lvt{\trans{\sbv l}\XS}
\def\Mvt{\trans{\sbv M}\XS}	\def\mvt{\trans{\sbv m}\XS}
\def\Nvt{\trans{\sbv N}\XS}	\def\nvt{\trans{\sbv n}\XS}
\def\Ovt{\trans{\sbv O}\XS}	\def\ovt{\trans{\sbv o}\XS}
\def\Pvt{\trans{\sbv P}\XS}	\def\pvt{\trans{\sbv p}\XS}
\def\Qvt{\trans{\sbv Q}\XS}	\def\qvt{\trans{\sbv q}\XS}
\def\Rvt{\trans{\sbv R}\XS}	\def\rvt{\trans{\sbv r}\XS}
\def\Svt{\trans{\sbv S}\XS}	\def\svt{\trans{\sbv s}\XS}
\def\Tvt{\trans{\sbv T}\XS}	\def\tvt{\trans{\sbv t}\XS}
\def\Uvt{\trans{\sbv U}\XS}	\def\uvt{\trans{\sbv u}\XS}
\def\Vvt{\trans{\sbv V}\XS}	\def\vvt{\trans{\sbv v}\XS}
\def\Wvt{\trans{\sbv W}\XS}	\def\wvt{\trans{\sbv w}\XS}
\def\Xvt{\trans{\sbv X}\XS}	\def\xvt{\trans{\sbv x}\XS}
\def\Yvt{\trans{\sbv Y}\XS}	\def\yvt{\trans{\sbv y}\XS}
\def\Zvt{\trans{\sbv Z}\XS}	\def\zvt{\trans{\sbv z}\XS}

\def\alphab{{\sbm \alpha}\XS}		\def\Gammab{{\sbm \Gamma}\XS}
\def\betab{{\sbm \beta}\XS}		\def\Deltab{{\sbm \Delta}\XS}
\def\gammab{{\sbm \gamma}\XS} 		\def\Thetab{{\sbm \Theta}\XS}
\def\deltab{{\sbm \delta}\XS}		\def\Lambdab{{\sbm \Lambda}\XS}
\def\epsilonb{{\sbm \epsilon}\XS} 	\def\Xib{{\sbm \Xi}\XS}
\def\varepsilonb{{\sbm \varepsilon}\XS} \def\Pib{{\sbm \Pi}\XS}
\def\zetab{{\sbm \zeta}\XS}		\def\Sigmab{{\sbm \Sigma}\XS}
\def\etab{{\sbm \eta}\XS}		\def\Varsigmab{{\sbm \Varsigma}\XS}
\def\thetab{{\sbm \theta}\XS}		\def\Upsilonb{{\sbm \Upsilon}\XS}
\def\varthetab{{\sbm \vartheta}\XS}	\def\Phib{{\sbm \Phi}\XS}
\def\iotab{{\sbm \iota}\XS}		\def\Psib{{\sbm \Psi}\XS}
\def\kappab{{\sbm \kappa}\XS}		\def\Omegab{{\sbm \Omega}\XS}
\def\lambdab{{\sbm \lambda}\XS}
\def\mub{{\sbm \mu}\XS}
\def\nub{{\sbm \nu}\XS}
\def\xib{{\sbm \xi}\XS} 
\def\pib{{\sbm \pi}\XS}
\def\varpib{{\sbm \varpi}\XS}
\def\rhob{{\sbm \rho}\XS}
\def\varrhob{{\sbm \varrho}\XS}
\def\sigmab{{\sbm \sigma}\XS}
\def\varsigmab{{\sbm \varsigma}\XS}
\def\phib{{\sbm \phi}\XS}
\def\varphib{{\sbm \varphi}\XS}
\def\chib{{\sbm \chi}\XS}
\def\psib{{\sbm \psi}\XS} 
\def\omegab{{\sbm \omega}\XS} 
\def\taub{{\sbm \tau}\XS}
\def\upsilonb{{\sbm \upsilon}\XS}

\def\alphabt{\trans{\sbm \alpha}\XS}		 \def\Gammabt{\trans{\sbm \Gamma}\XS}
\def\betabt{\trans{\sbm \beta}\XS}		 \def\Deltabt{\trans{\sbm \Delta}\XS}
\def\gammabt{\trans{\sbm \gamma}\XS} 		 \def\Thetabt{\trans{\sbm \Theta}\XS}
\def\deltabt{\trans{\sbm \delta}\XS}		 \def\Lambdabt{\trans{\sbm \Lambda}\XS}
\def\epsilonbt{\trans{\sbm \epsilon}\XS} 	 \def\Xibt{\trans{\sbm \Xi}\XS}
\def\varepsilonbt{\trans{\sbm \varepsilon}\XS}	 \def\Pibt{\trans{\sbm \Pi}\XS}
\def\zetabt{\trans{\sbm \zeta}\XS}		 \def\Sigmabt{\trans{\sbm \Sigma}\XS}
\def\etabt{\trans{\sbm \eta}\XS}		 \def\Varsigmabt{\trans{\sbm \Varsigma}\XS}
\def\thetabt{\trans{\sbm \theta}\XS}		 \def\Phibt{\trans{\sbm \Phi}\XS}
\def\varthetabt{\trans{\sbm \vartheta}\XS}	 \def\Psibt{\trans{\sbm \Psi}\XS}
\def\iotabt{\trans{\sbm \iota}\XS}		 \def\Omegabt{\trans{\sbm \Omega}\XS}
\def\kappabt{\trans{\sbm \kappa}\XS}		 \def\Upsilonbt{\trans{\sbm \Upsilon}\XS}
\def\lambdabt{\trans{\sbm \lambda}\XS}
\def\mubt{\trans{\sbm \mu}\XS}
\def\nubt{\trans{\sbm \nu}\XS}
\def\xibt{\trans{\sbm \xi}\XS} 
\def\pibt{\trans{\sbm \pi}\XS}
\def\varpibt{\trans{\sbm \varpi}\XS}
\def\rhobt{\trans{\sbm \rho}\XS}
\def\varrhobt{\trans{\sbm \varrho}\XS}
\def\sigmabt{\trans{\sbm \sigma}\XS}
\def\varsigmabt{\trans{\sbm \varsigma}\XS}
\def\phibt{\trans{\sbm \phi}\XS}
\def\varphibt{\trans{\sbm \varphi}\XS}
\def\chibt{\trans{\sbm \chi}\XS}
\def\psibt{\trans{\sbm \psi}\XS} 
\def\omegabt{\trans{\sbm \omega}\XS} 
\def\taubt{\trans{\sbm \tau}\XS}
\def\upsilonbt{\trans{\sbm \upsilon}\XS}

\def\alphabp{{\sbmp \alpha}\XS} 	     \def\Gammabp{{\sbmp \Gamma}\XS}
\def\betabp{{\sbmp \beta}\XS}		     \def\Deltabp{{\sbmp \Delta}\XS}
\def\gammabp{{\sbmp \gamma}\XS} 	     \def\Thetabp{{\sbmp \Theta}\XS}
\def\deltabp{{\sbmp \delta}\XS} 	     \def\Lambdabp{{\sbmp \Lambda}\XS}
\def\epsilonbp{{\sbmp \epsilon}\XS} 	     \def\Xibp{{\sbmp \Xi}\XS}
\def\varepsilonbp{{\sbmp \varepsilon}\XS}    \def\Pibp{{\sbmp \Pi}\XS}
\def\zetabp{{\sbmp \zeta}\XS}		     \def\Sigmabp{{\sbmp \Sigma}\XS}
\def\etabp{{\sbmp \eta}\XS}		     \def\Varsigmabp{{\sbmp \Varsigma}\XS}
\def\thetabp{{\sbmp \theta}\XS} 	     \def\Phibp{{\sbmp \Phi}\XS}
\def\varthetabp{{\sbmp \vartheta}\XS}	     \def\Psibp{{\sbmp \Psi}\XS}
\def\iotabp{{\sbmp \iota}\XS}		     \def\Omegabp{{\sbmp \Omega}\XS}
\def\kappabp{{\sbmp \kappa}\XS} 	     \def\Upsilonbp{{\sbmp \Upsilon}\XS}
\def\lambdabp{{\sbmp \lambda}\XS}
\def\mubp{{\sbmp \mu}\XS}
\def\nubp{{\sbmp \nu}\XS}
\def\xibp{{\sbmp \xi}\XS} 
\def\pibp{{\sbmp \pi}\XS}
\def\varpibp{{\sbmp \varpi}\XS}
\def\rhobp{{\sbmp \rho}\XS}
\def\varrhobp{{\sbmp \varrho}\XS}
\def\sigmabp{{\sbmp \sigma}\XS}
\def\varsigmabp{{\sbmp \varsigma}\XS}
\def\phibp{{\sbmp \phi}\XS}
\def\varphibp{{\sbmp \varphi}\XS}
\def\chibp{{\sbmp \chi}\XS}
\def\psibp{{\sbmp \psi}\XS} 
\def\omegabp{{\sbmp \omega}\XS} 
\def\taubp{{\sbmp \tau}\XS}
\def\upsilonbp{{\sbmp \upsilon}\XS}

\def\zerob{{\sbm 0}\XS}   \def\zerobp{{\sbmp 0}\XS}   \def\zerobt{\trans{\sbm 0}\XS}
\def\unb{{\sbm 1}\XS}	  \def\unbp{{\sbmp 1}\XS}     \def\unbt{\trans{\sbm 1}\XS}
\def\deuxb{{\sbm 2}\XS}   \def\deuxbp{{\sbmp 2}\XS}   \def\deuxbt{\trans{\sbm 2}\XS}
\def\troisb{{\sbm 3}\XS}  \def\troisbp{{\sbmp 3}\XS}  \def\troisbt{\trans{\sbm 3}\XS}
\def\quatreb{{\sbm 4}\XS} \def\quatrebp{{\sbmp 4}\XS} \def\quatrebt{\trans{\sbm 4}\XS}
\def\cinqb{{\sbm 5}\XS}   \def\cinqbp{{\sbmp 5}\XS}   \def\cinqbt{\trans{\sbm 5}\XS}
\def\sixb{{\sbm 6}\XS}	  \def\sixbp{{\sbmp 6}\XS}    \def\sixbt{\trans{\sbm 6}\XS}
\def\septb{{\sbm 7}\XS}   \def\septbp{{\sbmp 7}\XS}   \def\septbt{\trans{\sbm 7}\XS}
\def\huitb{{\sbm 8}\XS}   \def\huitbp{{\sbmp 8}\XS}   \def\huitbt{\trans{\sbm 8}\XS}
\def\neufb{{\sbm 9}\XS}   \def\neufbp{{\sbmp 9}\XS}   \def\neufbt{\trans{\sbm 9}\XS}

\def\eC{\mbox{$\mathbb{C}$}\XS} 		\def\eCp{\mathbb{C}\XS}
\def\eE{\mbox{$\mathbb{E}$}\XS} 		\def\eEp{\mathbb{E}\XS}
\def\eN{\mbox{$\mathbb{N}$}\XS} 		\def\eNp{\mathbb{N}\XS}
\def\eR{\mbox{$\mathbb{R}$}\XS} 		\def\eRp{\mathbb{R}\XS}
\def\eZ{\mbox{$\mathbb{Z}$}\XS} 		\def\eZp{\mathbb{Z}\XS}

\def\pth#1{\left(#1\right)}
\def\acc#1{\left\{#1\right\}}
\def\cro#1{\left[#1\right]}
\def\bars#1{\left|#1\right|}
\def\norm#1{\left\|#1\right\|}
\def\pol{\mbox{\large(}}	% Parenthese ouvrante large
\def\pfl{\mbox{\large)}}	% Parenthese fermante large
\def\poL{\mbox{\Large(}}	% Parenthese ouvrante Large
\def\pfL{\mbox{\Large)}}	% Parenthese fermante Large
\def\aol{\mbox{\large\{}}	% Accolade ouvrante large
\def\afl{\mbox{\large\}}}	% Accolade fermante large
\def\aoL{\mbox{\Large\{}}	% Accolade ouvrante Large
\def\afL{\mbox{\Large\}}}	% Accolade fermante Large
\def\col{\mbox{\large[}}	% Crochet ouvrant large
\def\cfl{\mbox{\large]}}	% Crochet fermant large
\def\coL{\mbox{\Large[}}	% Crochet ouvrant Large
\def\cfL{\mbox{\Large]}}	% Crochet fermant Large
\def\nl{\mbox{\large$\|$}}	% Norme large
\def\nL{\mbox{\Large$\|$}}	% Norme Large

\def\T{^\tDp}		% idem, plus compact \trans{} \equiv {}\T

\def\+{^\dagger}	% Transpos'e conjugu'e, … eviter dans les tabbing !
\def\I{\,|\,}		% "sachant" bien espac\'e pour les formules
\def\i{\,;\,}		% "point virgule" bien espac\'e pour les formules
\def\e#1{.10^{#1}}	% Notation scientifique a la francaise
\def\V{,\kern-.05em}	% Virgule pour les nombres decimaux

\def\egdef{\stackrel{\Delta}{=}}
\def\nequiv{\not\kern-.05em\equiv}
\def\egal{\kern-.5em=\kern-.5em} % Moins d'espace autour de "="
\def\dans{\in\!}	% Trop d'espace apres "appartient a"
\def\pourtt{\forall\,}  % Pas assez d'espace
\def\wh#1{\widehat{#1}} % Sombrero

\def\diag{\mathop{\rm diag}}
\def\Diag{\mathop{\rm Diag}}
\def\tr{\mathop{\rm tr}}
\def\esp{\mathop{\rm E}}
\def\Esp#1{{\esp\cro{#1}}}
\def\var{\mathop{\rm var}}
\def\reel#1{\rm Re}

\def\Card{{\rm Card}}
\def\arg{\mathop{\rm arg}}
\def\argmax{\mathop{\rm arg\,max}}	% Mieux que \def\argmax{\arg\max}
\def\argmin{\mathop{\rm arg\,min}}	% car l'indice est reparti
\def\froc#1#2{{#1/#2}}			% Frac en toc
\def\demi{\frac{1}{2}}
\def\diff#1#2{{\frac{d#1}{d#2}}}	% D'apres enquete, le "d droit"
\def\dd{\,d}				% doit etre en italique (et toc) !
\def\derpar#1#2{{\frac{\partial #1}{\partial #2}}}
\def\parsec#1#2#3{{\frac{\partial^2 #1}{\partial #2\,\partial #3}}}
\def\parsecd#1#2{{\frac{\partial^2 #1}{{\partial #2}^2}}}
\def\intdouble{\int\kern-0.85em\int}

\def\IF{{\rm if~}}
\def\SI{{\rm si~}}
\def\ET{{\rm et~}}
\def\AND{{\rm and~}}
\def\OU{{\rm ou~}}
\def\OR{{\rm or~}}
\def\FOR{{\rm for~}}
\def\SC{{\rm s.c.~}}	% Sous la contrainte...
\def\SINON{{\rm sinon}}
\def\OTHERWISE{{\rm otherwise}}

\def\sh{\mathop{\rm sh}}                % sin hyperbolique en FR
\def\ch{\mathop{\rm ch}}                % cos hyperbolique en FR
\def\th{\mathop{\rm th}}                % th hyperbolique en FR
\def\coth{\mathop{\rm coth}}            % cot hyperbolique en FR
\def\div{\mathop{\rm div}}              % divergence
\def\rotv{\overrightarrow{\mathop{\rm rot}}}    % rotationnel avec fleche
\def\gradv{\overrightarrow{\mathop{\rm grad}}}  % gradient avec fleche

\def\rond#1{\overset{\kern-0.33em~_\circ}{#1}}
\def\rondit[#1]#2{\overset{\kern#1~_\circ}{#2}}

\def\bm#1{\mbox{\boldmath $#1$}}
\def\zerob{{\bm 0}}
\def\oneb{{\bm 1}}

\def\ab{{\bm a}}
\def\bb{{\bm b}}
\def\cb{{\bm c}}
\def\db{{\bm d}}
\def\eb{{\bm e}}
\def\fb{{\bm f}}
\def\gb{{\bm g}}
\def\hb{{\bm h}}
\def\ib{{\bm i}}
\def\jb{{\bm j}}
\def\kb{{\bm k}}
\def\lb{{\bm l}}
\def\mb{{\bm m}}
\def\nb{{\bm n}}
\def\ob{{\bm o}}
\def\pb{{\bm p}}
\def\qb{{\bm q}}
\def\rb{{\bm r}}
\def\sb{{\bm s}}
\def\tb{{\bm t}}
\def\ub{{\bm u}}
\def\vb{{\bm v}}
\def\wb{{\bm w}}
\def\xb{{\bm x}}
\def\yb{{\bm y}}
\def\zb{{\bm z}}

\def\Ab{{\bm A}}
\def\Bb{{\bm B}}
\def\Cb{{\bm C}}
\def\Db{{\bm D}}
\def\Eb{{\bm E}}
\def\Fb{{\bm F}}
\def\Gb{{\bm G}}
\def\Hb{{\bm H}}
\def\Ib{{\bm I}}
\def\Jb{{\bm J}}
\def\Kb{{\bm K}}
\def\Lb{{\bm L}}
\def\Mb{{\bm M}}
\def\Nb{{\bm N}}
\def\Ob{{\bm O}}
\def\Pb{{\bm P}}
\def\Qb{{\bm Q}}
\def\Rb{{\bm R}}
\def\Sb{{\bm S}}
\def\Tb{{\bm T}}
\def\Ub{{\bm U}}
\def\Vb{{\bm V}}
\def\Wb{{\bm W}}
\def\Xb{{\bm X}}
\def\Yb{{\bm Y}}
\def\Zb{{\bm Z}}

\def\alphab{\bm{\alpha}}
\def\betab{\bm{\beta}}
\def\deltab{\bm{\delta}}
\def\epsilonb{\bm{\epsilon}}
\def\gammab{\bm{\gamma}}
\def\omegab{\bm{\omega}}
\def\thetab{\bm{\theta}}
\def\xib{\bm{\xi}}
\def\lambdab{\bm{\lambda}}
\def\taub{\bm{\tau}}
\def\phib{\bm{\phi}}
\def\mub{\bm{\mu}}
\def\psib{\bm{\psi}}
\def\chib{\bm{\chi}}
\def\sigmab{\bm{\sigma}}

\def\Deltab{\bm{\Delta}}
\def\Lambdab{\bm{\Lambda}}
\def\Phib{\bm{\Phi}}
\def\Psib{\bm{\Psi}}
\def\Sigmab{\bm{\Sigma}}

\def\Ac{{\cal A}}
\def\Bc{{\cal B}}
\def\Cc{{\cal C}}
\def\Dc{{\cal D}}
\def\Ec{{\cal E}}
\def\Fc{{\cal F}}
\def\Gc{{\cal G}}
\def\Hc{{\cal H}}
\def\Ic{{\cal I}}
\def\Jc{{\cal J}}
\def\Kc{{\cal K}}
\def\Lc{{\cal L}}
\def\Mc{{\cal M}}
\def\Nc{{\cal N}}
\def\Oc{{\cal O}}
\def\Pc{{\cal P}}
\def\Qc{{\cal Q}}
\def\Rc{{\cal R}}
\def\Sc{{\cal S}}
\def\Tc{{\cal T}}
\def\Uc{{\cal U}}
\def\Vc{{\cal V}}
\def\Wc{{\cal W}}
\def\Xc{{\cal X}}
\def\Yc{{\cal Y}}
\def\Zc{{\cal Z}}

\def\wt#1{\widetilde{#1}}
\def\wh#1{\widehat{#1}}
\def\xh{\widehat{x}}
\def\thetah{\widehat{\theta}}
\def\betah{\widehat{\beta}}

\def\xbh{\widehat{\xb}}
\def\thetabh{\widehat{\thetab}}
\def\betabh{\widehat{\betab}}

\def\xbhk{\widehat{\xb}^{k}}
\def\thetahk{\widehat{\theta}^{k}}
\def\betahk{\widehat{\beta}^{k}}

\def\xbhkp{\widehat{\xb}^{k+1}}
\def\thetahkp{\widehat{\theta}^{k+1}}
\def\betahkp{\widehat{\beta}^{k+1}}

\def\thetabhk{\widehat{\thetab}^{k}}
\def\betabhk{\widehat{\betab}^{k}}

\def\thetabhkp{\widehat{\thetab}^{k+1}}
\def\betabhkp{\widehat{\betab}^{k+1}}

\def\thetamin{\theta_{\mbox{\tiny min}}}
\def\thetamax{\theta_{\mbox{\tiny max}}}
\def\betamin{\beta_{\mbox{\tiny min}}}
\def\betamax{\beta_{\mbox{\tiny max}}}

\def\ra{\rightarrow}
\def\la{\leftarrow}
\def\da{\downarrow}
\def\ua{\uparrow}

\def\Ra{\Rightarrow}
\def\La{\Leftarrow}
\def\Da{\Downarrow}
\def\Ua{\Uparrow}

\def\lra{\longrightarrow}
\def\lla{\longleftarrow}
\def\Lra{\Longrightarrow}
\def\Lla{\longleftarrow}

\def\lrarr{\leftrightarrow}
\def\Lrarr{\Leftrightarrow}
\def\udarr{\updownarrow}
\def\Uparr{\Updownarrow}

\def\d#1{\,\mbox{d}#1}
\def\dxdy{\d{x}\d{y}}
\def\dwxdwy{\d{\omega_x}\d{\omega_y}}
\def\dxdydz{\d{x}\d{y}\d{z}}

\def\disp#1{{\displaystyle #1}}
\def\diag#1{\mbox{diag}\left\{#1\right\}}

\def\Prob#1{\mbox{Pr}\left\{#1\right\}}
\def\var#1{\mbox{Var}\left\{#1\right\}}
\def\cov#1{\mbox{Cov}\left\{#1\right\}}
\def\corr#1{\mbox{Corr}\left\{#1\right\}}
\def\trace#1{\mbox{Tr}\left\{#1\right\}}
\def\rang#1{\mbox{rang}\left\{#1\right\}}
\def\det#1{\mbox{d\'et}\left\{#1\right\}}

\def\cosf{\cos \phi}
\def\sinf{\sin \phi}
\def\cost{\cos \theta}
\def\sint{\sin \theta}

\def\sgn{\mbox{sgn}}
\def\sinc{\mbox{sinc}}
\def\rect{\mbox{rect}}
\def\sincf#1{\mbox{sinc}\left(#1\right)}
\def\rectf#1{\mbox{rect}\left(#1\right)}
\def\trif#1{\mbox{tri}\left(#1\right)}
\def\xvec#1#2#3{\left\{#1_#2,\ldots,#1_#3\right\}}

\def\vx{\left[x_1,\ldots, x_n\right]^t}
\def\vz{\left[z_1,\ldots, z_n\right]^t}
\def\vw{\left[\omega_1,\ldots, \omega_n\right]^t}
\def\vxi{\left[\xi_1,\ldots, \xi_n\right]^t}
\def\iii{\int_{-\infty}^{+\infty}}
\def\izi{\int_{0}^{\infty}}
\def\izpi{\int_{0}^{\pi}}
\def\izdpi{\int_{0}^{2\pi}}
\def\intd{\int\kern-.8em\int}
\def\intt{\int\kern-.8em\int\kern-.8em\int}
\def\intg{\int\kern-1.1em\int}
\def\sumd{\mathop{\sum\sum}}

\def\sumi{\sum_{i=1}^{M}}
\def\sumj{\sum_{i=1}^{N}}
\def\sumk{\sum_{k=1}^{K}}
\def\sumn{\sum_{n=1}^{N}}
\def\summ{\sum_{m=1}^{M}}

\def\TA#1{{\cal A}\left\{ {#1} \right\}}
\def\TH#1{{\cal H}\left\{ {#1} \right\}}
\def\TP#1{{\cal P}\left\{ {#1} \right\}}
\def\TR#1{{\cal R}\left\{ {#1} \right\}}
\def\TRa#1{{\cal R}^{\dag}\left\{ {#1} \right\}}
\def\BR#1{{\cal B}\left\{ {#1} \right\}}
\def\TF#1{{\cal F}\left\{ {#1} \right\}}
\def\TFI#1{{\cal F}^{-1}\left\{ {#1} \right\}}
\def\TFn#1#2{{\cal F}_{#1}\left\{ {#2} \right\}}
\def\TFnI#1#2{{\cal F}_{#1}^{-1}\left\{ {#2} \right\}}
\def\Im#1{{\cal I}\mbox{m}\left(#1\right)}
\def\Ker#1{{\cal K}\mbox{er}\left(#1\right)}
\def\Imag#1{\mbox{Im}\left(#1\right)}
\def\Re#1{\mbox{Re}\left(#1\right)}
\def\expf#1{\exp\left[ {#1} \right]}

\def\dfdx#1#2{{\mbox{d} {#1}\over{\mbox{d} {#2}}}}
\def\dfdxd#1#2{{\mbox{d}^2 {#1}\over{\mbox{d} {#2}^2}}}
\def\dfdxt#1#2{{\mbox{d}^3 {#1}\over{\mbox{d} {#2}^3}}}
\def\dfdxn#1#2{{\mbox{d}^n {#1}\over{\mbox{d} {#2}^n}}}
\def\dfdxk#1#2{{\mbox{d}^k {#1}\over{\mbox{d} {#2}^k}}}

\def\dpdx#1#2{\frac{\partial {#1}}{\partial {#2}}}
\def\dpdxd#1#2{\frac{\partial^2 {#1}}{\partial {#2}^2}}
\def\dpdxdy#1#2#3{{{\partial ^2 {#1}}\over{\partial {#2} \partial {#3}}}}

\def\arg{\mbox{arg}}
\def\argmins#1#2{\mbox{arg}\min_{#1}\left\{{#2}\right\}}
\def\argmaxs#1#2{\mbox{arg}\max_{#1}\left\{{#2}\right\}}
\def\argmin#1#2{\mathop{\mbox{arg}\min}_{#1}\left\{{#2}\right\}}
\def\argmax#1#2{\mathop{\mbox{arg}\max}_{#1}\left\{{#2}\right\}}

\def\esp#1{\mbox{E}\left\{ #1 \right\}}
\def\espx#1#2{\mbox{E}_{#1}\left\{ #2 \right\}}

\def\wth#1{\widehat{\widetilde{\phantom{#1}}}\!\!\!\! #1}

\def\lrf{L_{r,\phi}}
\def\fw{\widehat{f}(\omegab)}
\def\fthwxi{\wth{f}(\Omega,\xib)}
\def\fthwfi{\wth{f}(\Omega,\phi)}
\def\ftrfi{\widetilde{f}(r,\phi)}

\def\fwxwy{\widehat{f}(\omega_x, \omega_y)}
\def\wxpwy{(\omega_x \, x + \omega_y \, y)}

\def\wtx{\omegab^t \cdot \xb}
\def\ejwtx{\exp\left[j \omegab^t \cdot \xb\right]}
\def\xitx{\xib^t \cdot \xb}

\def\ftrxi{\widetilde{f}(r,\xib)}
\def\ent{-\int p(x) \, \ln p(x) \d{x}}

\def\mean#1{\left< #1 \right>}
\def\slnhn{\sum_{n=1}^N \lambda_n h_n(\rb)}
\def\slngn{\sum_{n=1}^N \lambda_n g_n(\rb)}
\def\smngn{\sum_{n=1}^N \mu_n g_n(\rb)}
\def\slmhm{\sum_{m=1}^N \lambda_m h_m(\rb)}
\def\vlambda{\bm{\lambda} = [\lambda_1,\ldots,\lambda_n]}

\def\apriori{{\em a priori} }
\def\aposteriori{{\em a posteriori} }

\def\titre#1{\bcc{\Large\bf #1}\ecc}

\def\AMD{Ali Mohammad--Djafari}
\def\LSSa{Laboratoire des Signaux et Syst\`emes 
(CNRS--ESE--UPS) \\ 
\'Ecole Sup\'erieure d'\'Electricit\'e \\ 
Plateau de Moulon, 91192 Gif sur Yvette Cedex, France.}

\def\ME{maximum entropy}
\def\pdf{probability distribution function}
\def\lm{Lagrange multipliers}
\def\fix#1{\phi _#1(x)}
\def\fin{\fix n}
\def\fik{\fix k}
\def\fiz{\fix 0}
\def\sfinz{\sum_{n=0}^N \lambda_n \, \fin}
\def\sfinu{\sum_{n=1}^N \lambda_n \, \fin}
\def\bl{\bm{\lambda}}
\def\bd{\bm{\delta}}
\def\blz{\bl ^0}
\def\gnl{G _n(\bl)}
\def\gnlz{G _n(\blz)}
\def\un{n=1,\dots, N}
\def\nn{n=0,\dots, N}

\def\finn{\fin , \nn}
\def\esfinz{\exp\,\left[ -\sfinz \right] }
\def\esfinu{\exp\,\left[ -\sfinu \right] }
\def\esxm{\exp\,\left[ -\sum_{m=0}^N \lambda_m \, x^m \right] }
\def\efin{\esp \fin }
\def\zl{Z(\bl)}
\def\finxi{\phi _n(x_i)}
\def\snfinxi{\sum_{n=1}^N \lambda_n \finxi}
\def\esnfinxi{\exp \left[ - \snfinxi \right]}
\def\smfinxi{\sum_{i=1}^M \finxi}

\def\ejnw{\exp \left( -j n \omega_0 x \right) }
\def\eejnw{\mbox{E} \left\lbrace \ejnw \right\rbrace}

\def\signed#1{{\unskip\nobreak\hfil\penalty50\hskip2em\mbox{}
\nobreak\hfil\tt#1\parfillskip=0pt \finalhyphendemerits=0 \par}}

\def\uncatcodespecials{\def\do##1{\catcode`##1=12 }\dospecials}
\def\listing#1{\par\begingroup\setupverbatim\input#1 \endgroup}
\newcount\lineno
\def\setupverbatim{\tt \lineno=0
 \obeylines \uncatcodespecials \obeyspaces
 \everypar{\advance\lineno by1 \llap{\sevenrm\the\lineno\ \ }}}
{\obeyspaces\global\let =\ }

\def\defined{\stackrel{\mbox{def}}{=}}
\def\str{\stackrel}

\def\ER{\mbox{I\kern-.25em R}}
\def\EC{\mbox{C\kern-.8em C}}
\def\EZ{\mbox{Z\kern-.55em Z}}
\def\EN{\mbox{N\kern-.8em N}}

\def\singles{
 \abovedisplayskip 12pt plus 3pt minus 9pt
 \belowdisplayskip 12pt plus 3pt minus 9pt
 \abovedisplayshortskip 0pt plus 3pt
 \belowdisplayshortskip 7pt plus 3pt minus 4pt
 \baselineskip 14.4pt
 \lineskip 1pt
 \lineskiplimit 0pt}
\def\oneandhalf{
 \abovedisplayskip 18pt plus 3pt minus 9pt
 \belowdisplayskip 18pt plus 3pt minus 9pt
 \abovedisplayshortskip 0pt plus 3pt
 \belowdisplayshortskip 9.333pt plus 3pt
 \baselineskip 20pt
 \lineskip 2pt
 \lineskiplimit 1pt}

\def\double{
 \abovedisplayskip 24pt plus 3pt minus 9pt
 \belowdisplayskip 24pt plus 3pt minus 9pt
 \abovedisplayshortskip 0pt plus 3pt
 \belowdisplayshortskip 12pt plus 3pt
 \baselineskip 27pt
 \lineskip 3pt
 \lineskiplimit 2pt}

\def\dadb{\d{\alpha}\d{\beta}}

\def\ffbox#1{\fbox{\mbox{\vbox{#1}}}}

\def\rot{\mbox{rot}}
\def\case#1#2#3#4{
    \left\{
           \begin{array}{ll}
            {\displaystyle #1} & {\displaystyle #2} \cr 
            {\displaystyle #3} & {\displaystyle #4}
           \end{array}
    \right. }

\def\beqnarr#1&#2&#3\\#4&#5&#6\eeqnarr{
    \left\{
           \begin{array}{lcl}
            {\displaystyle #1} & #2 & {\displaystyle #3} \\ 
            {\displaystyle #4} & #5 & {\displaystyle #6} 
           \end{array}
    \right. }

\def\pyx{p(\yb|\xb)}
\def\pxy{p(\xb|\yb)}

\def\ie{{\em i.e.}}
\def\unsdpi{\left(\frac{1}{2\pi}\right)}
\def\unspi{\left(\frac{1}{\pi}\right)}
\def\up{\uppercase}
\def\zjm{z_{j-1}}
\def\zjp{z_{j+1}}
\def\fxyp{f(x,y)=\left\{
\barr{ll} 1 & (x,y)\in P\\ 0 & (x,y)\not\in P\earr
\right.}

\def\rem#1{}

\def\cov#1{\mbox{Cov}\cro{#1}}
\def\esp#1{\mbox{E}\cro{#1}}
\def\espx#1#2{\mbox{E}_{#1}\cro{#2}}

\def\xtheta#1{\left(\frac{x_{#1}}{\theta_{#1}}\right)}
\def\xthetap#1#2{\xtheta{#1}^{\frac{#2}{\gamma_{#1}}}}

\def\xthetak{\left(\frac{x_{k}}{\theta_{k}}\right)}
\def\xthetal{\left(\frac{x_{l}}{\theta_{l}}\right)}

\def\xthetalc{\left(\frac{x_l}{\theta_l}\right)^{c_l}}
\def\xthetakc{\left(\frac{x_k}{\theta_k}\right)^{c_k}}
\def\xthetajc{\left(\frac{x_j}{\theta_j}\right)^{c_j}}

\def\Xtheta#1{\left(\frac{X_{#1}}{\theta_{#1}}\right)}
\def\Xthetap#1#2{\Xtheta{#1}^{\frac{#2}{\gamma_{#1}}}}

\def\Xthetak{\left(\frac{X_{k}}{\theta_{k}}\right)}
\def\Xthetal{\left(\frac{X_{l}}{\theta_{l}}\right)}

\def\Xthetalc{\left(\frac{X_l}{\theta_l}\right)^{c_l}}
\def\Xthetakc{\left(\frac{X_k}{\theta_k}\right)^{c_k}}
\def\Xthetajc{\left(\frac{X_j}{\theta_j}\right)^{c_j}}

\def\xthetalc{\left(\frac{x_l}{\theta_l}\right)^{c_l}}
\def\xthetakc{\left(\frac{x_k}{\theta_k}\right)^{c_k}}
\def\xthetajc{\left(\frac{x_j}{\theta_j}\right)^{c_j}}

\def\Xthetalc{\left(\frac{X_l}{\theta_l}\right)^{c_l}}
\def\Xthetakc{\left(\frac{X_k}{\theta_k}\right)^{c_k}}
\def\Xthetajc{\left(\frac{X_j}{\theta_j}\right)^{c_j}}

\def\ppj{\left(1+\sum_{j=1}^{n} \xthetap{j}{1}\right)}
\def\ppjc{\left(1+\sum_{j=1}^{n} \xthetajc^{c_j}\right)}

\def\ppjX{\left(1+\sum_{j=1}^{n} \Xthetap{j}{1}\right)}
\def\ppjcX{\left(1+\sum_{j=1}^{n} \Xthetajc^{c_j}\right)}

\def\fx{f_{n}(\xb)}
\def\fX{f_{n}(\Xb)}

\def\fcl{\frac{\xthetap{l}{1}}{\ppj}}
\def\fck{\frac{\xthetap{k}{1}}{\ppj}}
\def\fclc{\frac{\xthetalc}{\ppjc}}

\def\fclX{\frac{\Xthetap{l}{1}}{\ppjX}}
\def\fckX{\frac{\Xthetap{k}{1}}{\ppjX}}
\def\fclcX{\frac{\Xthetalc}{\ppjcX}}

\def\fcllog{\frac{\xthetap{l}{1}\ln\xthetal}{\ppj}}
\def\fcllogd{\frac{\xthetap{l}{1}\ln^2\xthetal}{\ppj}}

\def\fcllogX{\frac{\Xthetap{l}{1}\ln\Xthetal}{\ppjX}}
\def\fcllogdX{\frac{\Xthetap{l}{1}\ln^2\Xthetal}{\ppjX}}

\def\fcd{\frac{\xthetap{l}{2}}{\ppj^2}}
\def\ktan{\quad k=1, \cdots,n}
\def\ltan{\quad l=1, \cdots,n}

\def\psil{\Psi_{r_l}(\alpha)}
\def\psik{\Psi_{r_k}(\alpha)}
\def\psikl{\Psi_{r_l r_k}(\alpha)}

\def\xmut{\pth{\frac{x_l-\mu_l}{\theta_l}}}
\def\xmutl{\pth{\frac{x_l-\mu_l}{\theta_l}}^{m_l}}
\def\xmutk{\pth{\frac{x_k-\mu_k}{\theta_k}}^{m_k}}

\def\Xmut{\pth{\frac{X_l-\mu_l}{\theta_l}}}
\def\Xmutl{\pth{\frac{X_l-\mu_l}{\theta_l}}^{m_l}}
\def\Xmutk{\pth{\frac{X_k-\mu_k}{\theta_k}}^{m_k}}

\def\fxx{f_{X_l,X_k}(x_l,x_k)}
\def\dxx{\d{x_l}\d{x_k}}

\def\cktk{\frac{c_k}{\theta_k}}
\def\cltl{\frac{c_l}{\theta_l}}

\def\mata{\left[
\barr{ccc}
\barr{c} I(\theta_l) \\ I(\theta_l,\gamma_l) \\
I(\gamma_l,\alpha) \earr
& \barr{c} I(\theta_l,\gamma_l) \\
I(\gamma_l) \\ I(\gamma_l,\alpha) \earr&
\barr{c}
I(\theta_l,\alpha)
\\ I(\gamma_l,\alpha) \\ I(\alpha) \earr \earr\right]}

\def\matb{\left[
\barr{ccc} \barr{c} 1 \\ 0 \\ 0 \earr & \barr{c} 0 \\ \gamma^2 \\
0 \earr & \barr{c} 0
\\ 0 \\ 1 \earr \earr\right]}

\def\matbc{\left[
\barr{ccc} \barr{c} a \\ b \\ c \earr & \barr{c} d \\ e \\
f \earr & \barr{c} g
\\ h \\ i \earr \earr\right]}

\begin{document}
\title{Information and Covariance Matrices for Multivariate Burr III and Logistic distributions}

\shorttitle{Information and Covariance Matrices}

\authornames{Yari \& Mohammad-Djafari}

\authorone[Iran University of Science and Technology]{Gholamhossein Yari}
% Affiliation is just the name of your university or institution
\addressone{Iran University of Science and Technology, 
Narmak, Tehran 16844, Iran.
\\ {\tt email: \quad yari@iust.ac.ir}\\}
% Your postal address goes here.

\authortwo[Laboratoire des Signaux et Syst\`emes ({\sc Cnrs,Sup\'elec,Ups)}]
{Ali Mohammad-Djafari}
% Affiliation is just the name of your university or institution
\addresstwo{Sup\'elec, Plateau de Moulon, 3 rue Joliot-Curie, 
91192 Gif-sur-Yvette, France.
{\tt email: \quad djafari@lss.supelec.fr}}
% Your postal address goes here.

\begin{abstract}
Main result of this paper is to derive the exact analytical
expressions of information and covariance matrices for multivariate 
Burr III and logistic distributions. 
These distributions arise as tractable parametric models in 
price and income distributions,
reliability, economics, populations growth and survival data.
We showed that all the calculations can be obtained from one main
moment multi dimensional integral whose expression is obtained
through some particular change of variables. 
Indeed, we consider that this calculus technique for improper integral 
has its own importance in applied probability calculus.
\end{abstract}
% insert the primary Maths Subject Classification number in the first bracket
         % and the secondary ams number(s) in the second bracket
         % e.g. \ams{60E20}{49G03;49F10}

%%% ----------------------------------------------------------------------
\newpage
\section{Introduction}
\label{s1}

In this paper the exact form of Fisher information matrices for
multivariate Burr III and logistic distributions is determined. 
It is well-known that the information matrix is a valuable tool for  
derivation of covariance matrix in the asymptotic distribution of 
maximum likelihood estimations (MLE). 
In the univariate case for Pareto (IV) and Burr XII distributions, 
the Fisher information matrix expressions are given  
by Brazauskas~\cite{Brazauskas2003}
 and Watkins~\cite{Watkins1996}.
 As discussed in Serfling~\cite{Serfling1980}, 
%section~\ref{s4},  
under suitable regularity conditions, the determinant of the 
asymptotic covariance matrix of (MLE) reaches an optimal lower 
bound for the volume of the spread ellipsoid of joint estimators.

The univariate logistic distribution has been studied rather
extensively and, in fact, many of its developments through the
years were motivated to the normal distribution 
(see for example the handbook of Balakrishnan~\cite{Balakrishnan1992}). 
However, work on multivariate logistic distribution has been rather skimpy
compared to the voluminous work that has been carried out on
bivariate and multivariate normal distributions 
(Gumbel~\cite{Gumbel1961}, Arnold~\cite{Arnold1992}, 
Johnson, Kotz and Balakrishnan~\cite{Johnson1994} and 
Malik and Abraham~\cite{Malik1973}). 
For a broad discussion of logistic models and diverse applications see
 Malik and Abraham~\cite{Malik1973}.

Burr III and Logistic distributions also arise as tractable parametric 
models in the context of actuarial science, reliability,
economics, price and income distributions 
(Dagum~\cite{Dagum1996}, Burr~\cite{Burr1942} and Burr~\cite{Burr1967a}). 

This paper is organized as follows: 
Multivariate Burr III and logistic distribution are introduced and
presented in section~\ref{s2}. 
Elements of the information and covariance matrix for multivariate 
Burr III distribution is derived in section~\ref{s3}. 
Elements of the information and
covariance matrix for multivariate logistic distribution is
derived in section~\ref{s4}. 
Conculusion is presented in section~\ref{conclusion}.
 Derivation of first and second derivatives of the log$-$density function of
 multivariate Burr III distribution and calculation of its  main moment
 integral are given in Appendices $A$ and $B$. 
Derivation of first and second derivatives of the log density of
 multivariate logistic distribution and calculation of its main moment
 integral are given in Appendices $C$ and $D$.

\section{Multivariate Burr III and logistic distributions}
\label{s2} 

The density function of the Burr III distribution is
\beq\label{eq1}
f_{X}(x)=\frac{{\alpha}c\pth{\frac{x-\mu}{\theta}}^{-\pth{c+1}}}
{\theta\pth{1+(\frac{x-\mu}{\theta})^{-c}}^{\alpha+1}}, 
\quad x>\mu, 
\eeq 
where $-\infty<\mu<+\infty$ is the location parameter,
$\theta>0$ is the scale parameter, $c>0$ is the shape parameter
and $\alpha>0$ is the shape parameter which
 characterizes the tail of the distribution.

The $n$-dimensional Burr III distribution is 
\beq
 \label{eq2}
f_{n}(\xb)=\pth{1+\sum_{j=1}^n
\pth{\frac{x_{j}-\mu_{j}}{\theta_{j}}}^{-c_{j}}}^{-(\alpha+n)}
\prod_{i=1}^n \frac{(\alpha+i-1)c_{i}}{\theta_i}
\pth{\frac{x_{i}-\mu_{i}}{\theta_{i}}}^{-({c_i}+1)},
\eeq 
where $\xb=[x_1,\cdots,x_n]$, $x_{i}>\mu_{i}$, $c_{i}>0$,
$-\infty<\mu_{i}<+\infty$ , $\alpha>0$,
 $\theta_{i}>0$ for $i=1,  \cdots, n$. 
One of the main properties of this distribution is that,
the joint density of any subset of the components of a
multivariate Burr III random vector is again of the form
$(\ref{eq2})$ ~\cite{Johnson1994}.

The density of the logistic distribution is 
\beq
 \label{eq3}
f_{X}(x)=\frac{\alpha}{\theta} e^{-(\frac {x-\mu}{\theta})}
{\pth{1+e^{-(\frac {x-\mu}{\theta})}}^{-(\alpha+1)}}, \quad x>\mu,
\eeq 
where $-\infty<\mu<+\infty$ is the location parameter,
$\theta>0$ is the scale parameter and $\alpha>0$ is the shape
parameter.

The $n$-dimensional logistic distribution is\beq
 \label{eq4}
f_{n}(\xb)=\pth{1+\sum_{j=1}^n e^{-\pth{
\frac{x_{j}-\mu_{j}}{\theta_{j}}}}}^{-(\alpha+n)}\prod_{i=1}^ n
\frac{(\alpha+i-1)}{\theta_i} e^{-\pth{
\frac{x_{i}-\mu_{i}}{\theta_{i}}}},  
\eeq 
where
$\xb=[x_1,\cdots,x_n]$, $x_{i}>\mu_{i}$, $\alpha>0$,
$-\infty<\mu_{i}<+\infty$ and $\theta_{i}>0$ for 
$i=1,\cdots, n$. 
The joint density of any subset of the components of a
multivariate logistic random vector is again of the form
$(\ref{eq4})$ ~\cite{Johnson1994}.

%\newpage
\section{Information Matrix for Multivariate Burr III}
\label{s3}

Suppose $X$ is a random vector with the probability
density function $f_\Theta(. )$ where
 $\Theta=(\theta_{1}, \theta_{2}, . . . , \theta_{K})$.
 The information matrix $I(\Theta)$ is the $K\times K$ matrix
 with elements
\beq
 \label{eq5}
I_{ij}(\Theta)
=-\esp{\dpdxdy{\ln f_{\Theta}(\Xb)}{\theta_i}{\theta_j}},
\quad i,j=1,\cdots K.
\eeq
 For the multivariate Burr III, we have 
$\Theta =(\mu_1,\cdots, \mu_n, \theta_1,\cdots,\theta_n, c_1,\cdots, c_n, \alpha)$.
 In order to make the multivariate Burr III distribution a regular family
 (in terms of maximum likelihood estimation), we assume that vector $\mu$ is known and,
 without loss of generality, equal to 0. In this case information matrix is $(2n+1)\times(2n+1)$.
 Thus, further treatment is based on the following multivariate density function
\beq
 \label{eq6}
f_{n}(\xb)=\pth{1+\sum_{j=1}^n\pth{\frac{x_{j}}{\theta_{j}}}^{-c_{j}}}^{-(\alpha+n)}\prod_{i=1}^
n \frac{(\alpha+i-1)c_{i}}{\theta_{i}}
\pth{\frac{x_{i}}{\theta_{i}}}^{-({c_{i}+1})}, \quad x_{i}>0. 
\eeq
The log-density function is: 
\beqn 
\label{eq7} 
\ln f_{n}(\xb) 
&=& \sum_{i=1}^n\cro{ \ln (\alpha+i-1) -\ln\theta_{i} + \ln c_{i}}
-\pth{c_{i}+1}\ln\pth{\frac{x_{i}}{\theta_{i}}} \nonumber \\
&&-\pth{\alpha+n}\ln\pth{1+\sum_{j=1}^n \pth{\frac{x_{j}}{\theta_{j}}}^{{-c_{j}}}}.
\eeqn

Since the information
matrix $I(\Theta)$ is symmetric, it is enough to find elements
  $I_{ij}(\Theta)$, where $1 \leq i \leq j \leq  2n+1$.
The first and second partial derivatives of the above expression
are given in the Appendix $A$. In order to determine the
information matrix and score functions, we need to find expressions of the generic terms such as 
\[
\esp{\ln \pth{1+\sum_{j=1}^n \pth{\frac{X_{j}}{\theta_{j}}}^{{-c_{j}}}}}
\qquad \mbox{and} \qquad 
\esp{\Xtheta{l}^{-c_{l}}\Xtheta{k}^{-c_{k}}} \quad
%\esp{\Xtheta{l}^{-2c_{l}}},\quad \esp{\ln \Xtheta{l}}
\]
and evaluation of the required orders partial derivatives of the
last expectation at the required points.

\rem{
%\[
%\label{eq17} %\frac{\partial}{\partial {r_{l}}}\xi({r_{l}}=0) =
%\esp{\ln \Xtheta{l}},\quad %\eeq
\beq
%\frac{\partial}{\partial{r_{l}}}\xi({r_{l}}= {-c_{l}})=
%\esp{\Xtheta{l}^{-c_{l}}\ln \Xtheta{l}}, \quad
  %\frac{\partial^2}{\partial r^2_{l}}\xi({r_{l}}={-c_{l}})=
%\esp{\Xtheta{l}^{-c_{l}}\ln^2\Xtheta{l}},\quad
%\frac{\partial^2}{\partial{r_{l}^2}}\xi({r_{l}}={-2c_{l}})=
%\esp{\Xtheta{l}^{-2c_{l}}\ln^2 \Xtheta{l}} ,\quad
%%\]
%\[
%\frac{\partial^2}{\partial {r_{l}^2}}\xi({r_{l}}={-2c_{l}})=
%\esp{\Xtheta{l}^{-2c_{l}}\ln \Xtheta{l}},\quad%,\nonumber \eeq
\beq
%\esp{\Xtheta{l}^{-c_{l}}\Xtheta{k}^{-c_{k}}\ln\Xtheta{k}}
%,\quad,\nonumber \eeq
\beqn
%\]
%\[
%\esp{\Xtheta{l}^{-c_{l}}\Xtheta{k}^{-c_{k}}\ln\Xtheta{l}\ln\Xtheta{k}}%,\nonumber
%\eeqn
%S\]
% \esp{(\frac{X_{l}}{\theta_{l}})^{r_l}},\quad
%\esp{(\frac{X_{l}}{\theta_{l}})^{r_l}(\frac{X_{k}}{\theta_{k}})^{r_k}},\quad
%\esp{(\frac{X_{l}}{\theta_{l}})^{r_l}\ln^2\Xtheta{l}} ,\quad
%\esp{(\frac{X_{l}}{\theta_{l}})^{r_l}\ln\Xtheta{l}} ,\quad
%\]
%\beqnx {\esp{\frac{{\Xtheta{l}^{r_{l}}} {\Xtheta{k}^{r_{k}}}
%\ln^{n_{2}}\Xthetak\ln^{n_{1}}\Xthetal}{\pth{1+\sum_{j=1}^n
%(\frac{X_{j}}{\theta_{j}})^{{-c_{j}}}}^{n_{3}}}}},\\
%\quad (r_{l}, r_{k}\leq 0)\in\ER,\quad n_{1}, n_{2} \in\EN
%\mbox{~and}
%,\quad  n_{3}\in\ER^+. 
\eeqnx
%\quad \esp{(\frac{X_{l}}{\theta_{l}})^{{-c_{l}}}}
%\quad
%,\quad\esp{(\frac{X_{l}}{\theta_{l}})^{{-2c_{l}}}},\quad
  % \esp{(\frac{X_{l}}{\theta_{l}})^{{-c_{l}}}(\frac{X_{k}}{\theta_{k}})^{{-c_{k}}}},
%\[
%\esp{\frac{(\frac{X_{l}}{\theta_{l}})^{{-c_{l}}}}
%{\pth{1+\sum_{j=1}^n(\frac{X_{j}}{\theta_{j}})^{{-c_{j}}}}}}
%,\quad
%\esp{\ln\Xtheta{l}},\quad\esp{(\frac{X_{l}}{\theta_{l}})^{{-2c_{l}}}\ln\Xtheta{l}},\quad
%\]
%\[
%\esp{(\frac{X_{l}}{\theta_{l}})^{{-c_{l}}}\ln^2\Xtheta{l}}
%,\quad\esp{(\frac{X_{l}}{\theta_{l}})^{{-2c_{l}}}\ln^2\Xtheta{l}}
%,\quad
%\]
%and the general terms %\beqnx {\esp{\frac{{\Xtheta{l}^{r_{l}}}
%{\Xtheta{k}^{r_{k}}
}

%\newpage
\subsection{Main strategy to obtain expressions of the expectations}

Derivation of these  expressions are based on the following strategy: 
first, we derive an analytical expression for the
following integral 
\beq 
\label{eq8} 
\esp{\prod_{i=1}^n
\Xtheta{i}^{r_i}}= \int_{0}^{+\infty} \cdots
\int_{0}^{+\infty}\prod_{i=1}^n \xtheta{i}^{r_i}f_{n}(\xb)\d{\xb},
\eeq 
and then, we show that all the other expressions can easily be
found from it. We consider this derivation as one of the
main contributions of this work. This derivation is given in the
Appendix $B$. The result is the following: 
\beqn 
\label{eq9}
\esp{\prod_{i=1}^n \Xtheta{i}^{r_i}}
&=& \int_{0}^{+\infty} \cdots
\int_{0}^{+\infty}\prod_{i=1}^n \xtheta{i}^{r_i}f_{n}(\xb)\d{\xb} \nonumber \\
&=& \frac{\Gamma(\alpha+\sum_{i=1}^n \frac{r_{i}}{c_{i}}) \prod_{i=1}^n
  \Gamma(1- \frac{r_{i}}{c_{i}})}{\Gamma(\alpha)}
  =\sum_{i=1}^n \frac{r_{i}}{c_{i}}<\alpha,\quad
  \frac{r_{i}}{c_{i}}<1, \qquad \qquad 
\eeqn 
where $\Gamma$ is the usual Gamma function and 
\[
\Gamma_{r_lr_k}\left(\alpha+\sum_{i=1}^n\frac{r_{i}}{c_{i}}\right)=
\dpdxdy{\Gamma\left(\alpha+\sum_{i=1}^n\frac{r_{i}}{c_{i}}\right)}{r_k}{r_l},
\quad 1\leq {l, k} \leq n,
\]
\[
\Psi^{(n)} (z)=\frac{d^n}{d z^n}
\pth{\frac{\Gamma'(z)}{\Gamma(z)}},\quad z>0,
\]
\[
\frac{\partial^{(m+n)}}{\partial r_l^{m}\partial r_k^{n}}
 \pth{\frac{\Gamma_{r_l r_k}(z)}{\Gamma(z)}}=
\Psi^{(m+n)}(z),\quad z >0
\]
and $n,m\in\EN$.
%n,m \geq 0$.

Specifically, we use digamma
$\Psi(z)=\Psi^{(.)}(z)$, trigamma $\Psi'(z)$ and
 $\Psi_{r_l r_k}(z)$ functions 
(Abramowitz~\cite{Abramowitz1972} and Brazauskas~\cite{Brazauskas2003}).
To confirm the regularity of ln$f_{n}(\xb)$ and evaluation of the
expected Fisher information matrix, we take expectations of the
first and second order
 partial derivatives of $(\ref{eq7})$.
All the other expressions can be derived from this main result.
Derivative with respect to $\alpha$, from the both sides of the
relation 
\beq 
\label{eq10} 
1=\int_{0}^{+\infty} f_{n}(\xb)
\d{\xb}, 
\eeq 
leads to 
\beq 
\label{eq11}
\esp{\ln\pth{1+\sum_{j=1}^n(\frac{X_{j}}{\theta_{j}})^{{-c_{j}}}}}=\sum_{i=1}^n
\frac{1}{\alpha+i-1}. 
\eeq 
From relation $(\ref{eq9})$, for a pair of $(l, k)$ we have 
\beq 
\label{eq12}
\varphi(r_l,r_k)=\esp{\Xtheta{l}^{r_l}\Xtheta{k}^{r_k}}=
\frac{\Gamma(\alpha+\frac{r_l}{c_l}+\frac{c_k}{c_k})\Gamma(-\frac{r_l}{c_l}+1)
\Gamma(-\frac{r_k}{c_k}+1)}{\Gamma(\alpha)}. 
\eeq
From relation $(\ref{eq12})$, ~at $r_k=0$ we obtain 
\beq 
\label{eq14}
\xi({r_l})= \esp{\Xtheta{l}^{r_l}}=
\frac{\Gamma(\alpha+\frac{r_{l}}{c_l})
\Gamma(-\frac{r_{l}}{c_l}+1)}{\Gamma(\alpha)}.
\eeq
Evaluating this expectation at ${r_{l}}={-c_{l}}$,
 ${r_{l}}={-2c_{l}}$ and the relation $(\ref{eq12})$ at
 $({r_{l}}={-c_{l}}, {r_{k}}={-c_{k}})$, 
we obtain
\beqnx
\esp{\Xtheta{l}^{-c_{l}}}
&=&\frac{1}{\alpha-1}, 
\\
\esp{\Xtheta{l}^{-2c_{l}}}
&=&\frac{2}{(\alpha-1)(\alpha-2)}, 
\\
\esp{\Xtheta{l}^{-c_{l}}\Xtheta{k}^{-c_{k}}}
&=&\frac{1}{(\alpha-1)(\alpha-2)}.
\eeqnx

Evaluating the required orders partial derivatives of
$(\ref{eq14})$ and $(\ref{eq12})$ at the required points,  
we have

\beqnx
\esp{\ln \Xtheta{l}}
&=& \frac{\cro{\Psi(\alpha)-\Gamma'(1)}}{c_{l}},
\\ 
\esp{\Xtheta{l}^{-c_{l}}\ln \Xtheta{l}}
&=& \frac{\cro{\Psi(\alpha-1)-\Gamma'(2)}}{c_{l}(\alpha-1)}, 
\\
\esp{\Xtheta{l}^{-c_{l}}\ln^2\Xtheta{l}}
&=&\frac{\cro{\Psi^2(\alpha-1)+\Psi'(\alpha-1)
-2\Psi(\alpha-1)\Gamma'(2)+\Gamma''(2)}}{{c_{l}}^2(\alpha-1)} ,
\\
\esp{\Xtheta{l}^{-2c_{l}}\ln^2 \Xtheta{l}}
&=&\frac{\cro{2\Psi^2(\alpha-2)+2\Psi'(\alpha-2)
-2\Psi(\alpha-2)\Gamma'(3)+\Gamma''(3)}}{{c_l}^2(\alpha-1)(\alpha-2)},
\\
\esp{\Xtheta{l}^{-2c_{l}}\ln\Xtheta{l}}
&=&\frac{\cro{2\Psi(\alpha-2)-\Gamma'(3)}}{{c_{l}}(\alpha-1)(\alpha-2)},
\\
\esp{\Xtheta{l}^{-c_{l}}\Xtheta{k}^{-c_{k}}\ln\Xtheta{k}}
&=&\frac{\cro{\Psi_{r_l}(\alpha-2)-\Gamma'(2)}}{{c_k}(\alpha-1)(\alpha-2)},
\\
\esp{\Xtheta{l}^{-c_{l}}\Xtheta{k}^{-c_{k}}\ln\Xtheta{l}\ln\Xtheta{k}}
&=&
\frac{\cro{-\Gamma'(2)(\Psi_{r_l}(\alpha-2)+\Psi_{r_k}(\alpha-2))}}{{{c_l}{c_k}
(\alpha-1)(\alpha-2)}}  \\
& & +\frac{\cro{(\Gamma'(2))^2
+\Psi_{{r_l}{r_k}}(\alpha-2)}}{{c_l}{c_k}(\alpha-1)(\alpha-2)}. 
\eeqnx

%writing the expressions of the expectations\beq
%\esp{\frac{(\frac{X_{l}}{\theta_{l}})^{{-c_{l}}}}
%{\pth{1+\sum_{j=1}^n(\frac{X_{j}}{\theta_{j}})^{{-c_{j}}}}}},\quad
%\esp{\frac{(\frac{X_{l}}{\theta_{l}})^{{-2c_{l}}}}
%%{\pth{1+\sum_{j=1}^n(\frac{X_{j}}{\theta_{j}})^{{-c_{j}}}}^2}},\quad
%\esp{\frac{(\frac{X_{l}}{\theta_{l}})^{{-2c_{l}}}\ln\Xtheta{l}}
%{\pth{1+\sum_{j=1}^n(\frac{X_{j}}{\theta_{j}})^{{-c_{j}}}}^2}},\nonumber\eeq
%\beq
%\esp{\frac{(\frac{X_{l}}{\theta_{l}})^{{-2c_{l}}}\ln^2\Xtheta{l}}
%{\pth{1+\sum_{j=1}^n(\frac{X_{j}}{\theta_{j}})^{{-c_{j}}}}^2}},\quad
%\esp{\frac{(\frac{X_{l}}{\theta_{l}})^{{-c_{l}}}
%(\frac{X_{k}}{\theta_{k}})^{{-c_{k}}}\ln\Xtheta{l}\ln\Xtheta{k}}
%{\pth{1+\sum_{j=1}^n(\frac{X_{j}}{\theta_{j}})^{{-c_{j}}}}^2}},\nonumber
%\eeq
%\beq\esp{\frac{(\frac{X_{l}}{\theta_{l}})^{{-c_{l}}}\ln\Xtheta{l}}
%{\pth{1+\sum_{j=1}^n(\frac{X_{j}}{\theta_{j}})^{{-c_{j}}}}}},\quad
%\esp{\frac{(\frac{X_{l}}{\theta_{l}})^{{-c_{l}}}\ln^2\Xtheta{l}}
%{\pth{1+\sum_{j=1}^n(\frac{X_{j}}{\theta_{j}})^{{-c_{j}}}}}},\quad
%\esp{\frac{(\frac{X_{l}}{\theta_{l}})^{{-c_{l}}}
%(\frac{X_{k}}{\theta_{k}})^{{-c_{k}}}\ln\Xtheta{k}}
%{\pth{1+\sum_{j=1}^n(\frac{X_{j}}{\theta_{j}})^{{-c_{j}}}}^2}},\nonumber
%\eeq as $\mbox{~E}_{\alpha}$ to emphasis the role of the parameter
  %$\alpha$ in $(\ref{eq6})$, it can easily be shown that, by using of the

%\newpage
From these equations 
%$(\ref{eq15})$, $(\ref{eq16})$ and $(\ref{eq17})$ 
with $\alpha$  replaced by $(\alpha +1)$ and $(\alpha +2)$ in $(\ref{eq6})$,  
we can show that

\beqnx
\esp{\frac{(\frac{X_{l}}{\theta_{l}})^{{-c_{l}}}}
{\pth{1+\sum_{j=1}^n(\frac{X_{j}}{\theta_{j}})^{{-c_{j}}}}}} 
&=&
\frac{\alpha}{\alpha+n}\espx{\alpha+1}{\Xtheta{l}^{-c_{l}}}
=\frac{1}{\alpha+n},
\\
\esp{\frac{(\frac{X_{l}}{\theta_{l}})^{{-2c_{l}}}}
{\pth{1+\sum_{j=1}^n(\frac{X_{j}}{\theta_{j}})^{{-c_{j}}}}^2}} 
&=&\frac{\alpha(\alpha+1)}{(\alpha+n)(\alpha+n+1)}
\espx{\alpha+2}{\Xtheta{l}^{-2c_{l}}}
\\ 
&=&\frac{2}{(\alpha+n)(\alpha+n+1)},  
\\
\esp{\frac{(\frac{X_{l}}{\theta_{l}})^{{-c_{l}}}\ln^2\Xtheta{l}}
{\pth{1+\sum_{j=1}^n(\frac{X_{j}}{\theta_{j}})^{{-c_{j}}}}}}
&=&\frac{\cro{\Psi^2(\alpha)+\Psi'(\alpha)
-2\Psi(\alpha)\Gamma'(2)+\Gamma''(2)}}{{c_{l}}^2(\alpha+n)},
\\ 
\esp{\frac{(\frac{X_{l}}{\theta_{l}})^{{-c_{l}}}\ln\Xtheta{l}}
{\pth{1+\sum_{j=1}^n(\frac{X_{j}}{\theta_{j}})^{{-c_{j}}}}}}
&=&\frac{\cro{
 \Psi(\alpha)-\Gamma'(2)}}{c_{l}(\alpha+n)}, %\nonumber 
\\ 
\esp{\frac{(\frac{X_{l}}{\theta_{l}})^{{-2c_{l}}}\ln\Xtheta{l}}
{\pth{1+\sum_{j=1}^n(\frac{X_{j}}{\theta_{j}})^{{-c_{j}}}}^2}}
&=&\frac{\cro{2\Psi(\alpha)-\Gamma'(3)}}{{c_l}
(\alpha+n)(\alpha+n+1)},  
\\ 
\esp{\frac{(\frac{X_{l}}{\theta_{l}})^{{-2c_{l}}}\ln^2\Xtheta{l}}
{\pth{1+\sum_{j=1}^n(\frac{X_{j}}{\theta_{j}})^{{-c_{j}}}}^2}}
&=&\frac{\cro{2\Psi^2(\alpha)+\Gamma''(3)-2\Gamma'(3)\Psi(\alpha)
+2\Psi'(\alpha)}}{{c_l^2} (\alpha+n)(\alpha+n+1)},
\\ 
\esp{\frac{(\frac{X_{l}}{\theta_{l}})^{{-c_{l}}}
(\frac{X_{k}}{\theta_{k}})^{{-c_{k}}}\ln\Xtheta{k}}
{\pth{1+\sum_{j=1}^n(\frac{X_{j}}{\theta_{j}})^{{-c_{j}}}}^2}}
&=&\frac{\cro{\Psi_{r_k}(\alpha)-\Gamma'(2)}}{{c_k}
(\alpha+n)(\alpha+n+1)} ,
\\
\esp{\frac{(\frac{X_{l}}{\theta_{l}})^{{-c_{l}}}
(\frac{X_{k}}{\theta_{k}})^{{-c_{k}}}\ln\Xtheta{l}\ln\Xtheta{k}}
{\pth{1+\sum_{j=1}^n(\frac{X_{j}}{\theta_{j}})^{{-c_{j}}}}^2}}
&=&\frac{\cro{\Psi_{r_l}{_{r_k}}(\alpha)-\Gamma'(2)\pth{\Psi_{r_l}(\alpha)
+\Psi_{r_k}(\alpha)}+(\Gamma'(2))^2}}{{{c_l}{{c_k}
(\alpha+n)(\alpha+n+1)}}}. 
\eeqnx

\medskip
\subsection{Expectations of the score functions}

The expectations of the first three derivations of the first order follow immediately from the
 corresponding results for their three corresponding parameters and we obtain:

\beq\nonumber 
\label{eq20}\esp{\dpdx{\ln\fX}{\alpha}}=\sum_{i=1}^n
\frac{1}{\alpha+i-1}
 -\esp{\ln \pth{1+\sum_{j=1}^n(\frac{X_{j}}{\theta_{j}})^{{-c_{j}}}}}=0,
\eeq

\beq\nonumber 
\esp{\dpdx{\ln \fX}{\theta_l}}
=\frac{c_{l}}{\theta_{l}}
-\pth{\frac{(\alpha+n)c_{l}}{\theta_{l}}}
 \esp{\frac{(\frac{X_{l}}{\theta_{l}})^{{-c_{l}}}}
{\pth{1+\sum_{j=1}^n(\frac{X_{j}}{\theta_{j}})^{{-c_{j}}}}}}=0,
\eeq
 
\beq\nonumber  
 \esp{\dpdx{\ln\fX}{c_l}}= \frac{1}{c_l}-
\esp{\ln\Xtheta{l}} +\pth{\alpha+n}
 \esp{\frac{(\frac{X_{l}}{\theta_{l}})^{{-c_{l}}}\ln\Xtheta{l}}
{\pth{1+\sum_{j=1}^n(\frac{X_{j}}{\theta_{j}})^{{-c_{j}}}}}}=0.
\eeq

\subsection{The expected Fisher information matrix}

Main strategy is again based on the integral $(\ref{eq9})$ which
is presented in the Appendix $B$. After some tedious algebric
simplifications, the following expressions can be obtained

\beqnx
I_{\xb}(\alpha)
&=&\sum_{i=1}^n \frac{1}{\pth{\alpha+i-1}^2}, 
\\ 
I_{\xb}(\theta_l,\alpha)
&=&\frac{c_l}{\theta_l\pth{\alpha+n}},
\qquad l=1,\cdots,n, 
\\ 
I_{\xb}(c_l,\alpha)
&=&-\frac{1}{c_l\pth{\alpha+n}}\cro{\Psi(\alpha)-\Gamma'(2)},
\qquad l=1,\cdots,n, 
\\ 
I_{\xb}(\theta_l)
&=&\frac{c_l^2(\alpha+n-1)}{\theta_l^2({\alpha+n+1})}, 
\qquad l=1,\cdots,n,
\\
I_{\xb}(c_l)
&=&\frac{1}{c_l^2}
\cro{1+\Gamma''(2)-2\Psi(\alpha)\Gamma'(2)+\Psi^2(\alpha)+\Psi'(\alpha)}
\qquad l=1,\cdots,n,
\\
&&
-\frac{2}{c_l^2(\alpha+n+1)}\cro{-\Gamma'(3)\Psi(\alpha)+\Psi^2(\alpha)
+\Psi'(\alpha)+\frac{1}{2}\Gamma''(3)}, 
%\\ && \hspace*{4cm} 
\quad l=1,\cdots,n, 
\\ 
I_{\xb}(\theta_l,\theta_k)
&=&-\frac{c_l c_k}{\theta_l\theta_k\pth{\alpha+n+1}}, 
\qquad k\neq l, 
\\ 
I_{\xb}(c_l,c_k)
&=&-\frac{\cro{\pth{\Gamma'(2)}^2-\Gamma'(2)\pth{\Psi_{r_k}(\alpha)
+\Psi_{r_l}(\alpha)}+\Psi_{r_l}{_{r_k}}(\alpha)}}{c_l c_k\pth{\alpha+n+1}} ,  
\qquad k\neq l, 
\\
I_{\xb}(\theta_l,c_k)
&=&\frac{\cro{-\Gamma'(2)+\Psi_{r_k}
(\alpha)}}{\theta_l c_k \pth{\alpha+n+1}}, 
\qquad k\neq l, 
\\
I_{\xb}(\theta_l,c_l)
&=&\frac{\cro{\Gamma'(2)-\Psi(\alpha)}}{\theta_l}
 +\frac{\cro{2\Psi(\alpha)-\Gamma'(3)}}{\theta_l\pth{\alpha+n+1}}, 
\qquad l=1,\cdots,n. 
\eeqnx

Thus the information matrix,
$I_{\mbox{~Burr III}}(\Theta)$, for the multivariate Burr III$
(0, {\thetab}, {\cb}, \alpha)$ distribution is
\beq\label{eq32}
I_{\mbox{~Burr III}}(\Theta)= \left[ \barr{ccc}
  \barr{c} I(\theta_l, \theta_k) \\ 
I(\theta_l,c_k) \\ 
I(\theta_l, \alpha) \earr
& \barr{c} I(\theta_l, c_k) \\  I(c_l,c_k) \\
I(c_l, \alpha)   \earr & \barr{c} I(\theta_l, \alpha) \\
I(c_l, \alpha) \\  I(\alpha)  \earr 
\earr \right]. 
\eeq

\subsection{Covariance matrix for multivariate Burr III}

Since the joint density of any subset of the components of a
multivariate Burr III random vector is again of the form
$(\ref{eq2})$ 
%~\cite{Johnson1994},  
we can calculate the expectation
\beqn \label{eq33}
&& \hspace*{-2cm} \esp{\Xmutl\Xmutk}
\nonumber\\
&=&
 \int_{0}^{\infty} \int_{0}^{\infty} {\xmutl \xmutk} \fxx \dxx \nonumber
\nonumber\\
&=&\frac{\Gamma(\alpha+\frac{m_{l}}{c_{l}}+\frac{m_{k}}{c_{k}}) \Gamma(1-\frac{m_{l}}{c_{l}})\Gamma(1-\frac{m_{k}}{c_{k}})}{\Gamma(\alpha)},
\nonumber\\
&& 1-\frac{m_{l}}{c_{l}}>0,\quad 1-\frac{m_{k}}{c_{k}}>0,
\quad
\alpha+\frac{m_{l}}{c_{l}}+\frac{m_{k}}{c_{k}}>0. 
\eeqn

Evaluating this expectation at ($m_{l}=1$, $m_{k}=0$),
($m_{l}=0$, $m_{k}=1$)
 ($m_{l}=1$, $m_{k}=1$) and ($m_{l}=m$, $m_{k}=0$) we obtain

\beqnx 
\esp{X_{l}}
&=&\mu_{l}+\frac{\theta_{l}}{\Gamma(\alpha)}[\Gamma(\alpha+\frac{1}{c_l})
\Gamma(1-\frac{1}{c_l})], 
\\ 
\esp{X_{k}}
&=&\mu_{k}+\frac{\theta_{k}}{\Gamma(\alpha)}
[\Gamma(\alpha+\frac{1}{c_k})\Gamma(1-\frac{1}{c_k})], 
\\ 
\esp{X_{l}X_{k}}
&=&\mu_{k}\esp{X_{l}}+\mu_{l}\esp{X_{k}}-\mu_{l}\mu_{k}
\\
&& +\frac{\theta_{l}\theta_{k}}{\Gamma(\alpha)}
[\Gamma(\alpha+\frac{1}{c_k}+\frac{1}{c_k})
\Gamma(1-\frac{1}{c_k})\Gamma(1-\frac{1}{c_l})], 
\\
\esp{X_{l}^m}
&=&\frac{\theta_{l}^m}{\Gamma(\alpha)}[\Gamma(\alpha+\frac{m}{c_l})
\Gamma( 1-\frac{m}{c_l})],
\\
\var{X_l}
&=&\frac{\theta_{l}^2}{\Gamma^2(\alpha)}
\cro{\Gamma(\alpha+\frac{2}{c_l}) \Gamma(1-\frac{2}{c_l})
\Gamma(\alpha)
-\Gamma^2(1-\frac{1}{c_l})\Gamma^2(\alpha+\frac{1}{c_l})}, 
\quad 1-\frac{2}{c_l}>0,
\\ \cov{X_{l},X_{k}}
&=&\frac{\theta_{l}\theta_{k}}{\Gamma^2(\alpha)}\Gamma(1-\frac{1}{c_l})
\Gamma(1-\frac{1}{c_k})[\Gamma(\alpha+\frac{1}{c_l}
+\frac{1}{c_k})\Gamma(\alpha)
-\Gamma(\alpha+\frac{1}{c_k})\Gamma(\alpha+\frac{1}{c_l})]
\\
&& \qquad 1\leq l<k\leq n ,\qquad k=2,\cdots,n. \nonumber 
\eeqnx

\section{Information Matrix for Logistic distribution}

\label{s4} For the multivariate logistic distribution, we have
$\Theta = (\mu_1,\cdots, \mu_n, \theta_1,\cdots, \theta_n, \alpha)$.
 In order to make the multivariate logistic distribution a regular family
 (in terms of maximum likelihood estimation), we assume that vector $\mu$ is known and,
 without loss of generality, equal to 0. In this case information matrix is $(n+1)\times(n+1)$.

Thus, further treatment is based on the following multivariate 
density function
\beq\label{eq40}
f_{n}(\xb)=\pth{1+\sum_{j=1}^n
e^{-\pth{\frac{x_{j}}{\theta_{j}}}}}^{-(\alpha+n)}
\prod_{i=1}^ n
\frac{\pth{\alpha+i-1}}{\theta_i} 
e^{-\pth{\frac{x_{i}}{\theta_{i}}}}.
\eeq
Thus, the log-density function is:

\beq 
\label{eq41} \ln f_{n}(\xb)= \sum_{i=1}^n\cro{\ln
(\alpha+i-1)-\ln\theta_{i}}- \sum_{i=1}^n
(\frac{x_{i}}{\theta_{i}}) -\pth{\alpha+n}\ln\pth{1+\sum_{j=1}^n
e^{-( \frac{x_{j}}{\theta_{j}})}}.
\eeq
Since the information matrix $I(\Theta)$ is symmetric it is enough to find elements
  $I_{ij}(\Theta)$, where $1 \leq i \leq j \leq  n+1$.
The first and second partial derivatives of the above expression
are given in the Appendix $C$. Looking at these expressions, we
see that to determine the expression of the
 information matrix and score functions,  we need to find
  the following expectations
\[
\esp{\ln\pth{1+\sum_{j=1}^ne^{-\pth{\frac{X_{j}}{\theta_{j}}}}}}
\quad \mbox{and} \quad 
\esp{\pth{{e^{-\pth{\frac{X_{l}}{\theta_{l}}}}}}^{r_l}
\pth{{e^{-\pth{\frac{X_{k}}{\theta_{k}}}}}}^{r_k}},
%\esp{\frac{\Xtheta{l}^2e^{-2(\frac{X_{l}}{\theta_{l}})}}
%{\pth{1+\sum_{j=1}^ne^{-(\frac{X_{j}}{\theta_{j}})}}^2}},\quad
\]
and evaluation of the required orders partial derivatives of the
last expectation at the required points.

%\esp{\frac{\Xtheta{l}e^{-(\frac{X_{l}}{\theta_{l}})}}
%{\pth{1+\sum_{j=1}^ne^{-(\frac{X_{j}}{\theta_{j}})}}}},\quad
%\]
%\[
%\esp{\frac{\Xtheta{l}^2e^{-(\frac{X_{l}}{\theta_{l}})}}
%{\pth{1+\sum_{j=1}^ne^{-(\frac{X_{j}}{\theta_{j}})}}}},\quad
%\esp{\frac{
%\Xtheta{k}\Xtheta{l}e^{-(\frac{X_{l}}{\theta_{l}})}e^{-(\frac{X_{k}}{\theta_{k}})}}
%{\pth{1+\sum_{j=1}^ne^{-(\frac{X_{j}}{\theta_{j}})}}^2}}.
%\]
%and the general terms\beqnx \esp{\frac{\Xtheta{l}^{n_{1}}
%\Xtheta{k}^{n_{2}}{\pth{e^{-(\frac{X_{l}}{\theta_{l}})}}^{r_l}}{\pth{e^{-(\frac{X_{k}}
%{\theta_{k}})}}^{r_k}}}{\pth{1+\sum_{j=1}^
%ne^{-\pth{\frac{X_{j}}{\theta_{j}}}}}^{n_3}}},\\
%\quad (r_{l}, r_{k})\in\ER, n_{1}, n_{2} \in\EN^+
%\mbox{~and}\quad  n_{3}\in\ER^+. \eeqnx

\subsection{Main strategy to obtain expressions of the expectations}

Derivation of these  expressions are based on the following
strategy: first, we derive an analytical expression for the
following integral 
\beq\label{eq42} 
\esp{\prod_{i=1}^n\pth{
{e^{-\pth{\frac{X_{i}}{\theta_{i}}}}}}^{r_i}}= \int_{0}^{+\infty}
\cdots \int_{0}^{+\infty}\prod_{i=1}^n
\pth{{e^{-\pth{\frac{x_{i}}{\theta_{i}}}}}}^{r_i}f_{n}(\xb)\d{\xb},
\eeq 
and then, we show that all the other expressions can be
found from this easily. This derivation is given in the Appendix
$D$. The result is the following:
\beqn \label{eq43}
\esp{\prod_{i=1}^n\pth{{e^{-\pth{\frac{X_{i}}{\theta_{i}}}}}}^{r_i}}
&=&\int_{0}^{+\infty} \cdots
\int_{0}^{+\infty}{\prod_{i=1}^n
\pth{{e^{-\pth{\frac{x_{i}}{\theta_{i}}}}}}^{r_i}}
f_{n}(\xb)\d{\xb}
\nonumber \\
&=&\frac{\Gamma(\alpha-\sum_{i=1}^n{r_{i}}) \prod_{i=1}^n
  \Gamma(1+ {r_{i}})}{\Gamma(\alpha)}, \nonumber\\ 
&&\hspace*{2.5cm} \sum_{i=1}^n {r_{i}}<\alpha,\quad{r_{i}}>-1. 
\eeqn
Taking of derivative with respect to $\alpha$, from the both
sides of the relation 
\beq\label{eq44} 
1=\int_{0}^{+\infty} f_{n}(\xb) \d{\xb}, 
\eeq 
leads us to 
\beq\label{eq45}
\esp{\ln\pth{1+\sum_{j=1}^ne^{-\pth{\frac{X_{j}}{\theta_{j}}}}}}
=\sum_{i=1}^n \frac{1}{\alpha+i-1}. 
\eeq 
From relation $(\ref{eq43})$, for a pair of $(l, k)$ we have 
\beq\label{eq46}
\varphi(r_l,r_k)=\esp{\pth{{e^{-\pth{\frac{X_{l}}{\theta_{l}}}}}}^{r_l}
\pth{{e^{-\pth{\frac{X_{k}}{\theta_{k}}}}}}^{r_k}}=
\frac{\Gamma(\alpha-{r_l}-{r_k})\Gamma({r_k}+1)\Gamma({r_l}+1)}
{\Gamma(\alpha)}. 
\eeq 
From relation $(\ref{eq46})$, ~at $r_k=0$ we obtain 
\beq\label{eq47}
\esp{\pth{{e^{-\pth{\frac{X_{l}}{\theta_{l}}}}}}^{r_l}}=
\frac{\Gamma(\alpha-{r_l})\Gamma({r_l}+1)} {\Gamma(\alpha)}, 
\eeq
and evaluating this expectation at $r_{l}=1$, we obtain
\beq\label{eq48}
\esp{\pth{{e^{-\pth{\frac{X_{l}}{\theta_{l}}}}}}}=\frac{1}{\alpha-1} .
\eeq
Differentiating first and second order of $(\ref{eq47})$ with
respect to $r_{l}$ and replacing for $r_{l}=0$, $r_{l}=1$ and
$r_{l}=2$, we obtain the following relations:

\beqnx
\esp{\pth{\frac{X_l}{\theta_l}}}
&=&\Psi(\alpha)-\Gamma'(1), 
\\ 
\esp{\Xtheta{l}\pth{{e^{-\pth{\frac{X_{l}}{\theta_{l}}}}}}}
&=& \cro{\frac{\Psi(\alpha-1)-\Gamma'(2)}{(\alpha-1)}},
\\
\esp{\Xtheta{l}^2\pth{{e^{-\pth{\frac{X_{l}}{\theta_{l}}}}}}}
&=&\cro{\frac{\Psi^2(\alpha-1)-2\Gamma'(2) \Psi(\alpha-1)+\Psi'(\alpha-1)+\Gamma''(2)}{(\alpha-1)}},
\\
\esp{\Xtheta{l}^2e^{-2(\frac{X_{l}}{\theta_{l}})}}
&=&\frac{\cro{\Gamma''(3)-2\Gamma'(3)\Psi(\alpha-2)+2\Psi^2(\alpha-2)
+2\Psi'(\alpha-2)}}{(\alpha-1)(\alpha-2)}.
\eeqnx

From relation $(\ref{eq46})$,
\beqn \label{eq52}
 \frac{\partial}{\partial r_l\partial
r_k}\varphi(r_l=1,r_k=1)
&=& \esp{{\Xtheta{l}}
{\Xtheta{k}}\pth{e^{-\pth{\frac{X_{l}}{\theta_{l}}}}}
{\pth{e^{-\pth{\frac{X_{k}}{\theta_{k}}}}}}}
\nonumber\\
&=&\frac{\cro{\Gamma'(2)\pth{\Gamma'(2)-\Psi_{r_k}(\alpha-2)-\Psi_{r_l}(\alpha-2)}}}{(\alpha-1)(\alpha-2)}
\nonumber\\
&&+\frac{\Psi_{{r_l}{r_k}}(\alpha-2)}{(\alpha-1)(\alpha-2)}.
\eeqn

%Writing the expressions of the expectations\[
%\esp{\frac{\pth{e^{-(\frac{X_{l}}{\theta_{l}})}}}
%{\pth{1+\sum_{j=1}^ne^{-(\frac{X_{j}}{\theta_{j}})}}}},\quad
%Using $(\ref{eq3})$ with $\alpha$ replaced by  $(\alpha +1)$,
%we now obtain an expression for the last expectation as
%\esp{\frac{\Xtheta{l}e^{-(\frac{X_{l}}{\theta_{l}})}}
%{\pth{1+\sum_{j=1}^ne^{-(\frac{X_{j}}{\theta_{j}})}}}},\quad
%\esp{\frac{\Xtheta{l}^2e^{-(\frac{X_{l}}{\theta_{l}})}}
%{\pth{1+\sum_{j=1}^ne^{-(\frac{X_{j}}{\theta_{j}})}}}},
%\]
%\[
%\esp{\frac{\Xtheta{l}^2e^{-2(\frac{X_{l}}{\theta_{l}})}}
%{\pth{1+\sum_{j=1}^ne^{-(\frac{X_{j}}{\theta_{j}})}}^2}},\quad
%%\esp{\frac{
%%\Xtheta{k}\Xtheta{l}e^{-(\frac{X_{l}}{\theta_{l}})}e^{-(\frac{X_{k}}{\theta_{k}})}}
%{\pth{1+\sum_{j=1}^ne^{-(\frac{X_{j}}{\theta_{j}})}}^2}}.
%\]
 %as
%$\mbox{~E}_{\alpha}$ to emphasis the role of the parameter
  %$\alpha$ in $(\ref{eq40})$, it can easily be shown that by Using $(\ref{eq40})$

With $\alpha$ replaced by  $(\alpha +1)$ and $(\alpha +2)$ in $(\ref{eq40})$,
we obtain

\beqnx\label{eq53}
\esp{\frac{\pth{e^{\Xtheta{l}}}}{\pth{1+\sum_{j=1}^n e^{\Xtheta{j}}}}} 
&=&\frac{\alpha}{\alpha+n}
\espx{\alpha+1}{e^{\Xtheta{l}}}
=\frac{1}{\alpha+n},
\\
\esp{\frac{\Xtheta{l} e^{\Xtheta{l}}}{\pth{1+\sum_{j=1}^n e^{\Xtheta{j}}}}}
&=&\frac{\alpha}{\alpha+n}
\espx{\alpha+1}{\pth{\frac{X_{l}}{\theta_{l}}}{e^{-(\frac{X_{l}}{\theta_{l}})}}}
=\frac{\Psi(\alpha)-\Gamma'(2)}{\alpha+n},
\\
\esp{\frac{\Xtheta{l}^2 e^{\Xtheta{l}}}{\pth{1+\sum_{j=1}^n e^{\Xtheta{j}}}}}
&=&\frac{\Psi^2(\alpha)+\Psi'(\alpha)
+\Gamma''(2)-2\Gamma'(2)\Psi(\alpha)}{\alpha+n}, 
\\ 
\esp{\frac{\Xtheta{l}^2 e^{-2\Xtheta{l}}}{\pth{1+\sum_{j=1}^n e^{\Xtheta{j}}}}}
&=&\frac{\cro{2\Psi^2(\alpha)-2\Gamma'(3)\Psi(\alpha)+2\Psi'(\alpha)+\Gamma''(3)}}{{(\alpha+n)(\alpha+n+1)}},
\\ 
\esp{\frac{\Xtheta{k} \Xtheta{l} e^{-(\frac{X_{l}}{\theta_{l}})}
e^{-(\frac{X_{k}}{\theta_{k}})}}
{\pth{1+\sum_{j=1}^n e^{\Xtheta{j}}}^2}}
&=&\frac{\Gamma'(2)[{\Gamma'(2)-\Psi_{r_k}(\alpha)
-\Psi_{r_l}(\alpha)}]+\Psi_{{r_l}{r_k}}(\alpha)}
{(\alpha+n)(\alpha+n+1)}. 
\eeqnx

\subsection{Expectations of the score functions}

The expectations of the first two derivations of the first order
follow immediately from the
 corresponding results for their two corresponding parameters and we obtain:
\beq\nonumber
\esp{\dpdx{\ln\fX}{\alpha}}=\sum_{i=1}^n
\frac{1}{\alpha+i-1}
 -\esp{\ln \pth{1+\sum_{j=1}^ne^{-(\frac{X_{j}}{\theta_{j}})}}}=0,
\eeq 
\beq\nonumber 
\esp{\dpdx{\ln \fX}{\theta_l}}
=-\frac{1}{\theta_{1}}+\frac{1}{\theta_{l}}
\cro{\Psi(\alpha)-\Gamma'(1)}-\frac{1}{\theta_{l}}
\cro{\Psi(\alpha)-\Gamma'(2)}=0. 
\eeq

\subsection{The expected Fisher information matrix}

Main strategy is again based on the integral $(\ref{eq42})$ which
is presented in the Appendix $D$. 
Again after some tedious algebric
simplifications, the following expressions can be obtained

\beqnx
I_{\xb}(\alpha)
&=& \sum_{i=1}^n \frac{1}{\pth{\alpha+i-1}^2}, 
\\ 
I_{\xb}(\theta_l,\alpha)
&=& \frac{1}{\theta_l\pth{\alpha+n}}
\cro{\Psi(\alpha)-\Gamma'(2)},\quad l=1,\cdots,n, 
\eeqnx
\beqnx
I_{\xb}(\theta_l)
&=&\frac{(\alpha+n-1)}{\theta_l^2(\alpha+n+1)}
 \cro{\Psi^2(\alpha)-2\Gamma'(2)\Psi(\alpha)+\Gamma''(2)+\Psi'(\alpha)}\\
&&+\frac{1}{\theta_l^2}
-2\cro{\frac{\Gamma'(2)-\Psi(\alpha)}{\theta_l^2(\alpha+n+1)}},
\quad l=1,\cdots,n,  
\\ 
I_{\xb}(\theta_l,\theta_k)
&=&-\cro{\frac{\Gamma'(2)[\Gamma'(2)-\Psi_{r_k}(\alpha)
-\Psi_{r_l}(\alpha)]+\Psi_{r_l}{_{r_k}}(\alpha)}{\theta_l\theta_k(\alpha+n+1)}},\quad\quad k\neq l.
\eeqnx

Thus the information matrix, 
$I_{\mbox{ML}}(\Theta)$, for the multivariate logistic 
$(0, {\thetab},\alpha)$ distribution is 
\beq 
I_{\mbox{ML}}(\Theta)=
\left[ \barr{cc}
  \barr{c} I(\theta_l, \theta_k) \\ I(\theta_l, \alpha) \earr
&\barr{c} I(\theta_l, \alpha) \\ I(\alpha) \earr 
\earr \right].
\eeq

\subsection{Covariance matrix for multivariate Logistic }

Since the joint density of any subset of the components of a
multivariate logistic
 random vector is again multivariate logistic distribution$(\ref{eq4})$,
 we can use the relation $(\ref{eq46})$ with 
$\frac{\partial}{\partial r_l\partial r_k} \varphi(r_l=0,r_k=0)$ 
and obtain

\beqnx \label{eq64}
\esp{{X_l}{X_k}}
&=&{\theta_l}{\theta_k}\cro{\pth{\Gamma'(1)}^2-\Gamma'(1)\pth{\Psi_{r_k}(\alpha)
-\Psi_{r_l}(\alpha)}+\Psi_{r_l}{_{r_k}}(\alpha)},\quad k\neq l,
\\
\esp{X_l}
&=&\theta_l\cro{\Psi(\alpha)-\Gamma'(1)},\quad l=1,\cdots,n, 
\\
\esp{X_k}
&=&\theta_k\cro{\Psi(\alpha)-\Gamma'(1)},\quad k=1,\cdots,n. 
\eeqnx
From second order derivative of relation $(\ref{eq46})$, \ie, 
$\frac{\partial^2}{\partial r^2_l} \varphi(r_l=0,r_k=0)$
we have
\beqnx \label{eq67}
\esp{X^2_l}
&=&\theta^2_l\cro{(\Gamma''(1))-2\Gamma'(1)\Psi(\alpha)
+\Psi^2(\alpha)+\Psi'(\alpha)},\quad l=1,\cdots,n,
\\
\cov{X_{l},X_{k}}
&=&{\theta_l}{\theta_k}\cro{-\Gamma'(1)\pth{\Psi_{r_k}(\alpha)
+\Psi_{r_l}(\alpha)-2\Psi(\alpha)}+\Psi_{r_l}{_{r_k}}(\alpha)-\Psi^2(\alpha)},
\quad k\neq l,
\\
\var{X_l}
&=&\theta^2_{l}\cro{\Gamma''(1)-(\Gamma'(1))^2+\Psi'(\alpha)},
\quad l=1,\cdots,n. 
\eeqnx

\section{Conculusion}

\label{conclusion} In this paper we obtained the exact forms of
Fisher information and covariance matrices for multivariate Burr
III and multivariate logistic distributions. We showed that in
both distributions, all of the expectations can be obtained from
two main moment multi dimensional integrals which have been
considered and whose expression is obtained through some
particular change of variables. A short method of obtaining some
of the expectations as a function of $\alpha$ is used. To confirm
the regularity of the multivariate densities, we showed that the
expectations of the score functions are equal to $0$.

%\newpage
%\section*{Appendix A: Expressions of the derivatives}
\appendix
\section{Expressions of the derivatives}
\label{appendixA} \setcounter{equation}{0}~ In this Appendix, we
give in detail, the expressions for the first and second
derivatives of $\ln f_n(\xb)$ , where, $f_n(\xb)$ is the
multivariate Burr III density function $(\ref{eq6})$, which are
needed for obtaining the expression of the information matrix:

\beqnx
\dpdx{\ln \fx}{\alpha}
&=& \sum_{i=1}^n \frac{1}{\alpha+i-1} -
\ln{\pth{1+\sum_{j=1}^n(\frac{x_{j}}{\theta_{j}})^{-c_{j}}}},
\\
\dpdx{\ln \fx}{\theta_l}
&=& \frac{c_{l}}{\theta_{l}}
-\frac{(\alpha+n){c_l}}{\theta_{l}}
\frac{(\frac{x_{l}}{\theta_{l}})^
{-c_{l}}}{{\pth{1+\sum_{j=1}^n(\frac{x_{j}}
{\theta_{j}})^{-c_{j}}}}}, \quad l=1,\cdots,n,
\\
\dpdx{\ln \fx}{c_l}
&=&\frac{1}{c_l} -\ln\xtheta{l}
+\pth{\alpha+n}\frac{(\frac{x_{l}}{\theta_{l}})^
{-c_{l}}\ln\xtheta{l}}{{\pth{1+\sum_{j=1}^n(\frac{x_{j}}
{\theta_{j}})^{-c_{j}}}}} , \quad l=1,\cdots,n,
\eeqnx

\beqnx
\dpdxdy{\ln \fx}{\theta_l}{\alpha}
&=&-\frac{c_l}{\theta_l}\frac{(\frac{x_{l}}{\theta_{l}})^
{-c_{l}}}{{\pth{1+\sum_{j=1}^n(\frac{x_{j}}
{\theta_{j}})^{-c_{j}}}}}, \ltan,
\\
\dpdxdy{\ln \fx}{c_l}{\alpha}
&=&\frac{(\frac{x_{l}}{\theta_{l}})^
{-c_{l}}\ln\xtheta{l}}{{\pth{1+\sum_{j=1}^n(\frac{x_{j}}
{\theta_{j}})^{-c_{j}}}}}, \ltan,
\eeqnx

\beqnx
\dpdxd{\ln \fx}{\alpha}
&=&-\sum_{i=1}^n \frac{1}{(\alpha+i-1)^2},
\\
\dpdxd{\ln \fx}{\theta_l}
&=& -\frac{c_{l}}{\theta_l^2} +
\frac{(\alpha+n)(1-c_{l})c_{l}}
{\theta_l^2}\frac{(\frac{x_{l}}{\theta_{l}})^
{-c_{l}}}{{\pth{1+\sum_{j=1}^n(\frac{x_{j}}
{\theta_{j}})^{-c_{j}}}}}\\
&&
+\pth{\frac{(\alpha+n)c_{l}^2}{\theta_l^2}}\frac{(\frac{x_{l}}{\theta_{l}})^
{-2c_{l}}}{{\pth{1+\sum_{j=1}^n(\frac{x_{j}}
{\theta_{j}})^{-c_{j}}}}^2}, \ltan,
\\
\dpdxd{\ln \fx}{c_l}
&=&-\frac{1}{c_l^2}-\pth{\alpha+n}\frac{(\frac{x_{l}}{\theta_{l}})^
{-c_{l}}\ln^2\xthetal}{\pth{1+\sum_{j=1}^n(\frac{x_{j}}{\theta_{j}})^{-c_{j}}}}
+{\pth{\alpha+n}\frac{(\frac{x_{l}}{\theta_{l}})^
{-2c_{l}}\ln^2\xthetal}{\pth{1+\sum_{j=1}^n(\frac{x_{j}}
{\theta_{j}})^{-c_{j}}}^2}},
\eeqnx

\beqnx
\dpdxdy{\ln \fx}{\theta_k}{\theta_l}
&=& \pth{\frac{(\alpha+n)c_{l}
c_{k}}{\theta_l\theta_k}}
\frac{(\frac{x_{l}}{\theta_{l}})^{-c_{l}}(\frac{x_{k}}{\theta_{k}})^
{-c_{k}}}{\pth{1+\sum_{j=1}^n(\frac{x_{j}}
{\theta_{j}})^{-c_{j}}}^2}, \quad k\neq l, 
\\
\dpdxdy{\ln \fx}{c_k}{\theta_l}
&=&-\pth{\frac{(\alpha+n)c_{l}}{\theta_l}}
\frac{(\frac{x_{l}}{\theta_{l}})^{-c_{l}}(\frac{x_{k}}{\theta_{k}})^
{-c_{k}}\ln\xthetak}{\pth{1+\sum_{j=1}^n(\frac{x_{j}}
{\theta_{j}})^{-c_{j}}}^2}, \quad k\neq l,
\\
\dpdxdy{\ln \fx}{c_k}{c_l}
&=&\pth{\alpha+n}
\frac{(\frac{x_{l}}{\theta_{l}})^{-c_{l}}(\frac{x_{k}}{\theta_{k}})^
{-c_{k}}\ln\xthetak\ln\xthetal}{\pth{1+\sum_{j=1}^n(\frac{x_{j}}
{\theta_{j}})^{-c_{j}}}^2}, \quad k\neq l, 
\\
\dpdxdy{\ln \fx}{c_l}{\theta_l}
&=&
\frac{1}{\theta_l}
-\pth{\frac{\alpha+n}{\theta_l}}\frac{(\frac{x_{l}}{\theta_{l}})^
{-c_{l}}}{\pth{1+\sum_{j=1}^n(\frac{x_{j}}{\theta_{j}})^{-c_{j}}}}
\\
&&
+\pth{\frac{(\alpha+n)c_l}{\theta_l}}\frac{(\frac{x_{l}}{\theta_{l}})^
{-c_{l}}\ln\xthetal}{\pth{1+\sum_{j=1}^n(\frac{x_{j}}{\theta_{j}})^{-c_{j}}}}
\\
&&
-\pth{\frac{(\alpha+n)c_l}{\theta_l}}\frac{(\frac{x_{l}}{\theta_{l}})^
{-2c_{l}}\ln\xthetal}{{\pth{1+\sum_{j=1}^n(\frac{x_{j}}{\theta_{j}})^{-c_{j}}}}^2}
\ltan.
\eeqnx

%\newpage
\section{Expression of the main integral}
\label{appendixB}
\setcounter{equation}{0}

This Appendix gives one of the main results of this paper which
is the derivation of the expression of the following integral 
\beq\label{e1} 
\esp{\prod_{i=1}^n \Xtheta{i}^{r_i}}=
\int_{0}^{+\infty} \cdots \int_{0}^{+\infty}\prod_{i=1}^n
\xtheta{i}^{r_i}f_{n}(\xb)\d{\xb}, 
\eeq 
where, $f_{n}(\xb)$ is the multivariate Burr III density function (\ref{eq6}). 
This derivation is done in the following steps:

First consider the following one dimensional integral:
\beqnx 
C_{1}
&=&\int_{0}^{+\infty}\frac{\alpha c_{1}}{\theta_1}\xtheta{1}^{r_1}\xtheta{1}^{-(c_{1}+1)}
{\pth{1+\sum_{j=1}^n\xtheta{j}^{{-c_{j}}}}}^{-\pth{\alpha+n}}
\d{x}_{1} 
\nonumber
\\
&=&\int_{0}^{+\infty}\frac{\alpha
c_{1}}{\theta_1}\xtheta{1}^{r_1}\xtheta{1}^{-(c_{1}+1)}
{\pth{1+\sum_{j=2}^n\xtheta{j}^{{-c_{j}}}}}^{-\pth{\alpha+n}}
\nonumber
\\
&& \pth{1+\frac{\xtheta{1}^{-c_1}}
{1+\sum_{j=2}^n\xtheta{j}^{{-c_{j}}}}}^{-(\alpha+n)}
\d{x}_{1}. 
\eeqnx
Note that, goings from first line to second line
is just a factorizing and rewriting the last term of the
integrend. After many reflections on the links between  Burr
families and Gamma and Beta functions, we found that the following 
change of variable
\beq
\pth{1+\frac{\xtheta{1}^{-c_1}}
 {1+\sum_{j=2}^n\xtheta{j}^{{-c_{j}}}}}=\frac{1}{1-t},\quad 0<t<1,
\eeq 
simplifies this integral and guides us to the following result 
\beq
C_{1}=\frac{\alpha\Gamma(1-\frac{r_{1}}{c_1})\Gamma(\alpha+n-1+\frac{r_{1}}{c_{1}})}
{\Gamma(\alpha+n)}
\pth{{1+\sum_{j=2}^n\xtheta{j}^{-c_{j}}}}^{-(\alpha+n)-r_{1}+1}.
\eeq 
Then we consider the following similar expression: 
\beqnx
C_{2}
&=&\int_{0}^{+\infty}\frac{c_{2}\alpha(\alpha+1)\Gamma(1-\frac{r_{1}}{c_1})
\Gamma(\alpha+n-1+\frac{r_{1}}{c_{1}})}
{\theta_{2}\Gamma(\alpha+n)}
\\
&&\xtheta{2}^{r_{2}}
\xtheta{2}^{-(c_{2}+1)}
\pth{{1+\sum_{j=2}^n\xtheta{j}^{-c_{j}}}}^{-(\alpha+n)-\frac{r_{1}}{c_1}+1}
\d{x}_{2}, 
\\
&=&\int_{0}^{+\infty}\frac{c_{2}\alpha(\alpha+1)\Gamma(1-\frac{r_{1}}{c_1})
\Gamma(\alpha+n-1+\frac{r_{1}}{c_{1}})}
{\theta_{2}\Gamma(\alpha+n)}
\\
&&\xtheta{2}^{r_{2}}
\xtheta{2}^{-(c_{2}+1)}
\pth{{1+\sum_{j=2}^n\xtheta{j}^{-c_{j}}}}^{-(\alpha+n)-\frac{r_{1}}{c_1}+1}
\\
&& \pth{1+\frac{\xtheta{2}^{-c_2}}{\sum_{j=2}^n \xtheta{j}^{-c_j}}}^{-(\alpha+n)-\frac{r_{1}}{c_1}+1}
\d{x}_{2}, 
\eeqnx 
and again using the following change of variable:
\beq
 {\pth{1+\frac{\xtheta{2}^{-c_2}}
 {1+\sum_{j=3}^n\xtheta{j}^{{-c_{j}}}}}}=\frac{1}{1-t},
\eeq
we obtain: 
\beqn
C_{2}
&=&\frac{\alpha(\alpha+1)\Gamma(1-\frac{r_{1}}{c_1})
\Gamma(1-\frac{r_{2}}{c_2})\Gamma(\alpha+n-2+\frac{r_{1}}{c_{1}}
+\frac{r_{1}}{c_{1}})} {\Gamma(\alpha+n)}
\nonumber\\
&&
\pth{{1+\sum_{j=3}^n\xtheta{j}^{-c_{j}}}}^{-(\alpha+n)
-\frac{r_{1}}{c_{1}}-\frac{r_{2}}{c_{2}}+2}. 
\eeqn

Continuing this method, finally, we obtain the general
expression:

\beqn 
C_{n}
&=&\esp{\prod_{i=1}^n \Xtheta{i}^{r_i}}
\nonumber\\
&=&\frac{\Gamma(\alpha+\sum_{i=1}^n \frac{r_{i}}{c_i}) \prod_{i=1}^n
  \Gamma(1-\frac{r_{i}}{c_i})}{\Gamma(\alpha)},\quad  
  \sum_{i=1}^n \frac{r_{i}}{c_{i}}< \alpha,\quad 1-\frac{r_{i}}{c_{i} }>0. 
\eeqn

We may note that to simplify the lecture of the paper we did not give all
the details of these calculations.

%\newpage
\section{Expressions of the derivatives}
\label{appendixC} 
\setcounter{equation}{0} 
In this Appendix, we
give in detail the expressions for the first and second
derivatives of $\ln f_n(x)$, where, $f_{n}(\xb)$ is the
multivariate logistic density function (\ref{eq40}), which are
needed for obtaining the expression of the information matrix:

\beqnx 
\dpdx{\ln \fx}{\alpha}
&=& \sum_{i=1}^n \frac{1}{\alpha+i-1} -
\ln{\pth{1+\sum_{j=1}^ne^{-(\frac{x_{j}}{\theta_{j}})}}}, 
\\
\dpdx{\ln \fx}{\theta_l}
&=&- \frac{1}{\theta_{l}}
+\frac{x_l}{\theta^2_{l}}-\frac{(\alpha+n)}{\theta_{l}}
\frac{\pth{\frac{x_{l}}{\theta_{l}}}e^{-(\frac{x_{l}}{\theta_{l}})}}
{{\pth{1+\sum_{j=1}^ne^{-(\frac{x_{j}}{\theta_{j}})}}}} , \quad
l=1,\cdots,n, 
\\
\dpdxd{\ln \fx}{\alpha}
&=&-\sum_{i=1}^n \frac{1}{(\alpha+i-1)^2},
\\
\dpdxdy{\ln \fx}{\theta_l}{\alpha}
&=&-\frac{\pth{\frac{x_{l}}{\theta_{l}}}e^{-(\frac{x_{l}}{\theta_{l}})}}
{\theta_l{\pth{1+\sum_{j=1}^ne^{-(\frac{x_{j}}{\theta_{j}})}}}}, \ltan,
\\
\dpdxdy{\ln \fx}{\theta_k}{\theta_l}
&=&\frac{(\alpha+n)}{\theta_l\theta_k}
\frac{(\frac{x_{l}}{\theta_{l}})(\frac{x_{k}}{\theta_{k}})
e^{-(\frac{x_{l}}{\theta_{l}})}
e^{-(\frac{x_{k}}{\theta_{k}})}
}{\pth{1+\sum_{j=1}^ne^{-(\frac{x_{j}} {\theta_{j}})}}^2}, \quad
k\neq l, 
\\
\dpdxd{\ln \fx}{\theta_l}
&=&\frac{1}{\theta^2_l}-\frac{2x_l}{\theta^3_l} +\frac{2(\alpha+n)}
{\theta^2_l}\frac{\pth{\frac{x_{l}}{\theta_{l}}}e^{-(\frac{x_{l}}{\theta_{l}})}}
{{\pth{1+\sum_{j=1}^ne^{-(\frac{x_{j}}{\theta_{j}})}}}}
\\
&&
-\frac{(\alpha+n)}{\theta^2_l}\frac{\pth{\frac{x_{l}}{\theta_{l}}}^2
e^{-(\frac{x_{l}}{\theta_{l}})}}
{{\pth{1+\sum_{j=1}^ne^{-(\frac{x_{j}}{\theta_{j}})}}}}
+\frac{(\alpha+n)}{\theta^2_l}\frac{\pth{\frac{x_{l}}{\theta_{l}}}^2
e^{-2(\frac{x_{l}}{\theta_{l}})}}
{{\pth{1+\sum_{j=1}^ne^{-(\frac{x_{j}}{\theta_{j}})}}}^2}, 
\\ 
&& \hspace*{3cm} \ltan.
\eeqnx

%\newpage

\section{Expression of the main integral}
\label{appendixD} \setcounter{equation}{0}

This Appendix gives the second main result of this paper which is
the derivation of the expression of the following integral 
\beq
\esp{\prod_{i=1}^n\pth{
{e^{-\pth{\frac{X_{i}}{\theta_{i}}}}}}^{r_i}}= \int_{0}^{+\infty}
\cdots \int_{0}^{+\infty}\prod_{i=1}^n
\pth{{e^{-\pth{\frac{x_{i}}{\theta_{i}}}}}}^{r_i}f_{n}(\xb)\d{\xb},
\eeq 
where, $f_{n}(\xb)$ is
the multivariate logistic density function (\ref{eq40}).
 This derivation is done in the following steps:\\
 First consider the following one dimensional integral:
\beqn 
C_{n}
&=&\int_{0}^{+\infty}\frac{(\alpha+n-1)}
{\theta_n}\pth{e^{-(\frac{x_{n}}{\theta_{n}})}}^{r_n}e^{-(\frac{x_{n}}{\theta_{n}})}
{\pth{1+\sum_{j=1}^{n}e^{-(\frac{x_{j}}{\theta_{j}})}}}^{-(\alpha+n)}\d{x}_{n}
\nonumber\\
&=&\int_{0}^{+\infty}\frac{(\alpha+n-1)}
{\theta_n}\pth{e^{-(\frac{x_{n}}{\theta_{n}})}}^{r_n}e^{-(\frac{x_{n}}{\theta_{n}})}
{\pth{1+\sum_{j=1}^{n-1}e^{-(\frac{x_{j}}{\theta_{j}})}}}^{-(\alpha+n)}
\nonumber\\
&& \pth{1+\frac{e^{-(\frac{x_{n}}{\theta_{n}})}}
{1+\sum_{j=1}^{n-1}e^{-(\frac{x_{j}}{\theta_{j}})}}}^{-(\alpha+n)}\d{x}_{n}. 
\eeqn 
Note that, goings from first line to second line
is just a factorizing and rewriting the last term of the
integrend. 
After looking for the links between logistic
function  and Gamma and Beta functions, we found that the following
 change of variable
\beq 
\pth{1+\frac{e^{-(\frac{x_{n}}{\theta_{n}})}}
{1+\sum_{j=1}^{n-1}e^{-(\frac{x_{j}}{\theta_{j}})}}}
=\frac{1}{1-t},\quad 0<t<1, 
\eeq 
simplifies this integral and
guides us to the following result 
\beq
C_{n}=\frac{\Gamma(r_{n}+1)\Gamma(\alpha+n-r_{n}-1)}
{\Gamma(\alpha+n-1)}
\pth{{1+\sum_{j=1}^{n-1}e^{-(\frac{x_{j}}{\theta_{j}})}}}^{-(\alpha+n)+r_{n}+1}.
\eeq 
Then we consider the following similar expression: 
\beqn
C_{n-1}
&=&\frac{\Gamma(r_{n}+1)\Gamma(\alpha+n-r_{n}-1)}
{\Gamma(\alpha+n-1)} \int_{0}^{+\infty}\frac{(\alpha+n-2)}
{\theta_{n-1}}\pth{e^{-(\frac{x_{n-1}}{\theta_{n-1}})}}^{r_{n-1}}
\nonumber\\
&&
e^{-(\frac{x_{n-1}}{\theta_{n-1}})}\pth{{1+\sum_{j=1}^{n-1}e^{-(\frac{x_{j}}{\theta_{j}})}}}^
{-(\alpha+n)+r_{n}+1}\d{x}_{n-1} 
\nonumber\\
&&= \frac{\Gamma(r_{n}+1)\Gamma(\alpha+n-r_{n}-1)}
{\Gamma(\alpha+n-1)} \int_{0}^{+\infty}\frac{(\alpha+n-2)}
{\theta_{n-1}}\pth{e^{-(\frac{x_{n-1}}{\theta_{n-1}})}}^{r_{n-1}}
\nonumber\\
&&
e^{-(\frac{x_{n-1}}{\theta_{n-1}})}\pth{{1+\sum_{j=1}^{n-2}e^{-(\frac{x_{j}}{\theta_{j}})}}}^
{-(\alpha+n)+r_{n}+1}
\pth{1+\frac{e^{-(\frac{x_{n-1}}{\theta_{n-1}})}}
{1+\sum_{j=1}^{n-2}e^{-(\frac{x_{j}}{\theta_{j}})}}}^{-(\alpha+n)+r_{n}+1}
\d{x}_{n-1} 
\nonumber\\ 
\eeqn
and again using the following change of variable:
\[
 \pth{1+\frac{e^{-(\frac{x_{n-1}}{\theta_{n-1}})}}
{1+\sum_{j=1}^{n-2}e^{-(\frac{x_{j}}{\theta_{j}})}}}=\frac{1}{1-t},
\]
we obtain: 
\beqn
C_{n-1}
&=&\frac{\Gamma(r_{n-1}+1)\Gamma(r_{n}+1)\Gamma(\alpha+n-r_{n}-r_{n-1}-1)}
{\Gamma(\alpha+n-2)}
\nonumber\\
&&\pth{{1+\sum_{j=1}^{n-2}e^{-(\frac{x_{j}}{\theta_{j}})}}}^
{-(\alpha+n)+r_{n}+r_{n-1}+1}
\eeqn 

Continuing this method, finally, we obtain the general expression:
\beqn
C_{1}
&=&\esp{\prod_{i=1}^n\pth{{e^{-\pth{\frac{X_{i}}{\theta_{i}}}}}}^{r_i}}
\nonumber\\
&=&\frac{\Gamma(\alpha-\sum_{i=1}^n{r_{i}}) \prod_{i=1}^n
  \Gamma(1+ {r_{i}})}{\Gamma(\alpha)}, \nonumber
\quad \sum_{i=1}^n {r_{i}}<\alpha,\quad{r_{i}}>-1. 
\eeqn
We may note that to simplify the lecture of the paper we did not
give all the details of these calculations.

\newpage
\bibliographystyle{amsplain}
\bibliography{mybib}
\end{document}